\documentclass[twocolumn,showpacs,preprintnumbers,
amsmath,amssymb,aps,prd,nofootinbib,superscriptaddress,
eqsecnum]{revtex4}
\usepackage{graphicx}

\makeatletter
\renewcommand{\p@subsection}{}
\makeatother

\renewcommand{\thesection}{\arabic{section}}

\renewcommand{\theequation}{\arabic{section}.\arabic{equation}}

\begin{document}

\preprint{\vbox{\hbox{DPNU-05-18}\hbox{November, 2005}}}

\title{%
Dropping $\rho$ and $A_1$ Meson Masses
at Chiral Phase Transition\\
in the Generalized Hidden Local Symmetry~\footnote{%
 Main result of this work was presented in the talk
 ``Systematic perturbation theory based on generalized hidden
 local symmetry'' at Japan Physical Society meeting, March 24,
 2005, Noda, Japan.
}
}

\author{Masayasu Harada}
\affiliation{%
Department of Physics, Nagoya University,
Nagoya, 464-8602, Japan}
\author{Chihiro Sasaki}
\affiliation{%
Gesellschaft f\"{u}r Schwerionenforschung (GSI),
64291 Darmstadt, Germany
}

\begin{abstract}
We study the chiral symmetry restoration using the generalized
hidden local symmetry (GHLS) which incorporates the
rho and $A_1$ mesons as the gauge bosons of the GHLS and the pion
as the Nambu-Goldstone boson consistently with the chiral
symmetry of QCD.
We show that a set of parameter relations,
which 
ensures the first and second 
Weinberg sum rules, is invariant under the renormalization
group evolution.
Then, we found that the Weinberg sum rules
together with the
matching of the vector and axial-vector current correlators
inevitably leads to {\it the dropping masses of both rho and 
$A_1$ mesons} at the symmetry restoration point,
and that the mass ratio as well as the
mixing angle between the pion and $A_1$ meson 
flows into one of three fixed points.
\end{abstract}


\maketitle


\section{Introduction}
\label{sec:intro}

Changes of the hadron masses are indications
of the chiral symmetry restoration 
occurring in hot and/or dense QCD~\cite{rest}.
Dropping masses of hadrons following the Brown-Rho (BR)
scaling~\cite{BR-scaling} can be 
one of the most prominent candidates of the strong
signal of the melting of the quark condensate 
$\langle\bar{q}q\rangle$ which is
the order parameter of the spontaneous chiral symmetry
breaking.
Especially, 
the dropping of the $\rho$ meson mass
according to the BR scaling 
satisfactorily explained~\cite{LKB}
the enhancement of dielectron mass spectra below
the $\rho / \omega$ resonance observed at CERN SPS~\cite{ceres}.

The vector manifestation (VM)~\cite{HY:VM} is the
Wigner realization in which
the $\rho$ meson becomes massless degenerate with the
pion at the chiral phase transition point.
The VM is formulated in the effective field theory (EFT) based
on the hidden local symmetry (HLS)~\cite{BKUYY,BKY:PRep}.
In the HLS theory we can perform the systematic
chiral perturbation with the dynamical $\rho$ meson 
included~\cite{Georgi,HY:PLB,Tanabashi,HY:PRep}.
Furthermore, the matching to QCD \^^ a la Wilson 
combined with the renormalization group equations (RGEs)
gives several physical predictions 
in remarkable agreement with experiments~\cite{HY:WM,HY:PRep}.

The formulation of the VM was done also in hot 
matter~\cite{HS:VM} and in dense matter~\cite{HKR:VM},
and 
a compelling evidence of dropping mass recently comes
from the mass shift of the $\omega$ meson in nuclei measured by
the KEK-PS E325 Experiment~\cite{KEK-PS} and
the CBELSA/TAPS Collaboration~\cite{trnka}
and also from that of
the $\rho$ meson observed in the STAR experiment~\cite{SB:STAR}.
Since the VM formulated in the HLS theory
provides
a theoretical description of the dropping $\rho$ mass
which is protected by the existence of the
fixed point (VM fixed point),
we can study several other phenomena associated with the
dropping $\rho$ by expanding the HLS theory around the 
VM fixed point:
Large violation of the vector dominance
of the pion electromagnetic form factor should occur 
near the VM fixed point~\cite{HS:VD},
which plays an important role~\cite{Brown:2005ka-kb} 
to explain the recent experimental data provided by
NA60~\cite{NA60};
The pion velocity near the restoration point is predicted as
$v_\pi(T_c) = 0.83$-$0.99$~\cite{PiV}, which seems to be 
consistent with value extracted~\cite{Cramer}
from the recent data from the STAR collaboration at 
RHIC~\cite{STAR:PiV}.

In the VM, it was assumed that the axial-vector and scalar mesons
are decoupled from the theory near the phase transition point.
However, the masses of these mesons may decrease following the
BR scaling.  
Actually, recently in Ref.~\cite{BLR},
it was proposed to extend the VM to 
include axial-vector mesons
for explaining the 
anomalous $\rho^0/\pi^-$ ratio measured in peripheral
collisions by STAR~\cite{STAR}.
There were several analyses with models including 
axial-vector mesons such as in Ref.~\cite{rhoA1model}.
These analyses are not based on the fixed point structure
and found no significant reduction of the masses of
axial-vector meson.
Then, it is desirable to construct an EFT
which includes the axial-vector meson as a dynamical degree of
freedom, and study 
whether a fixed point structure exists and it can realize
the light axial-vector meson.

There are several models which includes the axial-vector 
meson in addition to the pion and vector meson consistently
with the chiral symmetry of QCD such as the
``Massive Yang-Mills'' field method~\cite{MassiveYM},
the anti-symmetric tensor field method~\cite{anti:tensor}
and the model based on the generalized hidden local symmetry
(GHLS)~\cite{BKY:NPB,BFY:GHLS,BKY:PRep,Kaiser:1990yf}.
These models are equivalent~\cite{Kaiser:1990yf,equi}
at least tree-level on-shell amplitude is concerned.
However, there are differences in the off-shell amplitude since the
definitions of the off-shell fields are different in the 
models (see, e.g., Ref.~\cite{HY:PRep}).
Here we pick up the model based on the GHLS which is a natural
extension of the HLS to include the axial-vector meson.

In this paper, we first develop the chiral perturbation theory
(ChPT) with GHLS,
in which a systematic low-energy expansion is possible 
even including the axial-vector meson in addition to
the pseudoscalar and vector mesons 
as a dynamical degree of freedom.
Then, we make the matching of the vector and
axial-vector current correlators with those obtained by
the operator product
expansion (OPE) in the energy region higher than the
axial-vector meson mass to find that
the resultant set of the parameter relations satisfies the 
pole saturated forms of the first and second Weinberg
sum rules.
Based on the RGEs
in the Wilsonian sense obtained in the ChPT with GHLS, 
the set of the parameter relations
is shown to be stable against the renormalization group evolution.

We further study the fate of the axial-vector meson
near the chiral restoration point, and found that
the Weinberg sum rules together with 
the matching necessarily leads to the dropping 
masses of both vector and axial-vector mesons.
Interestingly, the ratio of masses of vector and axial-vector 
mesons as well as the mixing between the pseudoscalar and
axial-vector mesons flows into one of three fixed points:
They exhibit the VM--like, Ginzburg-Landau--like 
and Hybrid--like patterns of the chiral symmetry restoration.

This paper is organized as follows:

In section~\ref{sec:GHLS}, we give a brief review on
the GHLS.
Construction of the ChPT with GHLS is done
in section~\ref{sec:CHPT}.
In section~\ref{sec:WSR}, we make the matching
to derive a set of parameter relations satisfying the 
pole saturated forms of the first and second Weinberg
sum rules.
This is shown to be stable against the renormalization 
group evolution.
Section~\ref{sec:PS} is devoted to study the
phase structure of the GHLS with the Weinberg sum 
rules kept satisfied.
We show that
both $\rho$ and $A_1$ necessarily become massless
at the phase transition point,
and that the mass ratio flows into one of three fixed points.
In section~\ref{sec:CRM}, we discuss the relation of 
three classes of the fixed point
to the chiral representation mixing.
Finally in section~\ref{sec:SD}, 
we give a brief summary and discussions.
We show the quantization of the GHLS theory based on the
background field gauge in Appendix~\ref{app:BFG},
and several quantum corrections and RGEs in 
Appendix~\ref{app:QC}.


\section{Generalized Hidden Local Symmetry}
\label{sec:GHLS}

The generalized hidden local symmetry (GHLS) is an
extention of the hidden local symmetry (HLS),
in which the axial-vector mesons as well as the vector mesons
are introduced as the gauge bosons of the GHLS, in addition 
to the pseudoscalar mesons as the Nambu-Goldstone bosons
associated with the spontaneous chiral symmetry breaking.
In this section, we briefly review the GHLS 
following Refs.~\cite{BKY:NPB,BKY:PRep,BFY:GHLS}.

\subsection{Lagrangian}
\label{ssec:lagrangian}

The GHLS Lagrangian is based on
the $G_{\rm{global}} \times G_{\rm{local}}$ symmetry,
where $G_{\rm global}=[SU(N_f)_L \times SU(N_f)_R]_{\rm global}$ 
is the chiral symmetry and 
$G_{\rm local}=[SU(N_f)_L \times SU(N_f)_R]_{\rm local}$ 
is the GHLS.
The whole symmetry $G_{\rm global}\times G_{\rm local}$
is spontaneously broken down to a flavor diagonal $SU(N_f)_V$.
The basic quantities are
the GHLS gauge bosons $L_\mu$ and $R_\mu$ and 
three matrix valued variables $\xi_L$, $\xi_R$
and $\xi_M$ which are introduced as
\begin{equation}
U = \xi_L^\dagger \xi_M \xi_R\,,
\end{equation}
where $N_f \times N_f$ special-unitary matrix $U$ is
a basic ingredient of the chiral perturbation theory (ChPT)
~\cite{Gasser:1983yg,Gasser:1984ux}.
The transformation property of $U$ under the chiral symmetry
is given by
\begin{equation}
U \to g_L U g_R^\dagger\,,
\end{equation}
where $g_L$ and $g_R$ are the elements of the chiral symmetry,
$g_{L,R} \in [SU(N_f)_{L,R}]_{\rm global}$.
The variables $\xi$s transform as
\begin{eqnarray}
&&
\xi_{L,R} \to h_{L,R}\cdot\xi_{L,R}\cdot g_{L,R}^\dagger\,,
\nonumber\\
&&
\xi_M \to h_L\cdot\xi_M\cdot h_R^\dagger\,,
\end{eqnarray}
with $h_{L,R} \in [SU(N_f)_{L,R}]_{\rm local}$.
The GHLS gauge fields $L_\mu$ and $R_\mu$ transform as
\begin{eqnarray}
&&
L_\mu \to ih_L\partial h_L^\dagger + h_L L_\mu h_L^\dagger\,,
\nonumber\\
&&
R_\mu \to ih_R\partial h_R^\dagger + h_R R_\mu h_R^\dagger\,.
\end{eqnarray}
The covariant derivatives of $\xi_{L,R,M}$ are given by
\begin{eqnarray}
&&
D_\mu \xi_L 
 = \partial_\mu\xi_L - iL_\mu\xi_L + i\xi_L{\cal{L}}_\mu\,,
\nonumber\\
&&
D_\mu \xi_R 
 = \partial_\mu\xi_R - iR_\mu\xi_R + i\xi_R{\cal{R}}_\mu\,,
\nonumber\\
&&
D_\mu \xi_M 
 = \partial_\mu\xi_M - iL_\mu\xi_M + i\xi_M R_\mu\,,
\end{eqnarray}
where ${\cal{L}}_\mu$ and ${\cal{R}}_\mu$ are the external
gauge fields introduced by gauging $G_{\rm{global}}$ symmetry.

The fundamental objects are the Maurer-Cartan 1-forms
defined by
\begin{eqnarray}
&&
\hat{\alpha}_{L,R}^\mu = D^\mu\xi_{L,R}\cdot\xi_{L,R}^\dagger /i\,,
\nonumber\\
&&
\hat{\alpha}_M^\mu = D^\mu\xi_M\cdot\xi_M^\dagger /(2i)\,,
\end{eqnarray}
which transform as
\begin{eqnarray}
&&
\hat{\alpha}_{L,R}^\mu 
 \to h_{L,R}\hat{\alpha}_{L,R}^\mu h_{L,R}^\dagger\,,
\nonumber\\
&&
\hat{\alpha}_M^\mu
 \to h_L \hat{\alpha}_M^\mu h_L^\dagger\,.
\label{alpha}
\end{eqnarray}
There are four independent terms, with the lowest derivatives,
invariant under $G_{\rm global}\times G_{\rm local}$:
\begin{eqnarray}
&&
{\cal L}_V 
 = F^2 \mbox{tr}\bigl[ \hat{\alpha}_{\parallel\mu}
   \hat{\alpha}_\parallel^\mu \bigr]\,,
\nonumber\\
&&
{\cal L}_A 
 = F^2 \mbox{tr}\bigl[ \hat{\alpha}_{\perp\mu}
   \hat{\alpha}_\perp^\mu \bigr]\,,
\nonumber\\
&&
{\cal L}_M 
 = F^2 \mbox{tr}\bigl[ \hat{\alpha}_{M\mu}
   \hat{\alpha}_M^\mu \bigr]\,,
\nonumber\\
&&
{\cal L}_\pi 
 = F^2 \mbox{tr}\bigl[ \bigl( \hat{\alpha}_{\perp\mu}
   {}+ \hat{\alpha}_{M\mu} \bigr)
   \bigl( \hat{\alpha}_{\perp}^\mu
   {}+ \hat{\alpha}_{M}^\mu \bigr)\bigr]\,,
\label{lag a-d}
\end{eqnarray}
where $F$ is the parameter
carrying the mass dimension $1$~\footnote{%
 In Ref.~\cite{BKY:NPB,BKY:PRep,BFY:GHLS}, each term has 
 the pion decay constant $F_\pi^2$ as the coefficient
 by taking the proper normalization.
 In this paper, however, we introduce $F$ just as a
 parameter which carries mass dimension $1$.
 In the latter section, we will define $F_\pi$ as
 the coupling strength to the broken current by dissolving
 the $\pi$-$A_1$ mixing.
}
and $\hat{\alpha}_{\parallel,\perp}$ are defined as
\begin{eqnarray}
&&
\hat{\alpha}_{\parallel,\perp}^\mu
 = \bigl( \xi_M\hat{\alpha}_R^\mu\xi_M^\dagger 
  {}\pm \hat{\alpha}_L^\mu \bigr)/2\,.
\end{eqnarray}

Another building block is the gauge field strength 
of the GHLS defined by
\begin{eqnarray}
&&
L_{\mu\nu}
 = \partial_\mu L_\nu - \partial_\nu L_\mu
  {}- i\bigl[ L_\mu, L_\nu \bigr]\,,
\nonumber\\
&&
R_{\mu\nu}
 = \partial_\mu R_\nu - \partial_\nu R_\mu
  {}- i\bigl[ R_\mu, R_\nu \bigr]\,.
\end{eqnarray}
{}From these field strengths, the kinetic term of the gauge bosons 
are given by 
\begin{equation}
{\cal L}_{\rm kin}(L_\mu,R_\mu)
 = {}- \frac{1}{4g^2}\mbox{tr}\bigl[ L_{\mu\nu}L^{\mu\nu}
   {}+ R_{\mu\nu}R^{\mu\nu} \bigr]\,,
\label{lag kin}
\end{equation}
with $g$ being the gauge coupling constant of the GHLS.
Note that the parity invariance requires that there is only 
one gauge coupling.

By combining the four terms in Eq.~(\ref{lag a-d}) together with
the kinetic term of the gauge fields in Eq.~(\ref{lag kin}),
the GHLS Lagrangian is given by
\begin{equation}
{\cal L} = a{\cal L}_V + b{\cal L}_A + c{\cal L}_M
 {}+ d{\cal L}_\pi 
 {}+ {\cal L}_{\rm kin}(L_\mu,R_\mu)\,,
\label{lag p^2}
\end{equation}
where $a$, $b$, $c$ and $d$ are dimensionless parameters 
to be determined by the underlying QCD.


\subsection{Particle Identification}
\label{ssec:PID}

The symmetry breaking pattern of the GHLS is given as
\begin{eqnarray}
&&
\left[SU(N_f)_L \times SU(N_f)_R\right]_{\rm local}
\nonumber\\
&& \quad
\times 
\left[SU(N_f)_L \times SU(N_f)_R\right]_{\rm global}
\nonumber\\
&& \qquad
\ \rightarrow\ 
SU(N_f)_V
\ ,
\end{eqnarray}
which generates $ 3 \times (N_f^2 -1) $ massless
Nambu-Goldstone (NG) bosons.
$ 2 \times (N_f^2 -1) $ of the NG bosons are absorbed into the
gauge bosons of the GHLS to give masses through the Higgs
mechanism.
$(N_f^2-1)$ NG bosons remain as the massless particles, which 
we identify with the pseudoscalar mesons $\pi$
(pion and its flavor partners).
On the other hand,
we identify the gauge bosons 
$V_\mu = (R_\mu + L_\mu)/2$ with the vector mesons 
denoted by $\rho$ 
($\rho$ meson and its flavor partners) and
$A_\mu = (R_\mu - L_\mu)/2$ with the axial-vector
mesons denoted as $A_1$ ($a_1$ meson and its flavor partners).

In the following, we specify the $\pi$ and
would-be NG bosons absorbed into $\rho$ and $A_1$
by parameterizing $\xi_{L,R,M}$ as
\begin{eqnarray}
&&
\xi_{R} = e^{i(\phi_\sigma + \phi_\perp)}\,,
\nonumber\\
&&
\xi_{L} = e^{i(\phi_\sigma - \phi_\perp)}\,,
\nonumber\\
&&
\xi_M = e^{2i\phi_p}\,.
\end{eqnarray}
Three 1-forms are expanded into
\begin{eqnarray}
&&
\hat{\alpha}_\parallel^\mu
 = \partial^\mu\phi_\sigma
  {}- V^\mu + {\cal V}^\mu + \cdots\,,
\nonumber\\
&&
\hat{\alpha}_\perp^\mu
 = \partial^\mu\phi_\perp
  {}- A^\mu + {\cal A}^\mu + \cdots\,,
\nonumber\\
&&
\hat{\alpha}_M^\mu
 = \partial^\mu\phi_p + A^\mu + \cdots\,,
\label{alpha_expand}
\end{eqnarray}
where the vector and axial-vector external gauge fields
${\mathcal V}_\mu$ and ${\mathcal A}_\mu$ are defined as
\begin{equation}
{\mathcal V}_\mu = \frac{1}{2} \left(
  {\mathcal R}_\mu + {\mathcal L}_\mu
\right)
\ , \quad
{\mathcal A}_\mu = \frac{1}{2} \left(
  {\mathcal R}_\mu - {\mathcal L}_\mu
\right)
\ .
\end{equation}
The $a {\mathcal L}_V$ term in the Lagrangian
is expressed as
\begin{equation}
a {\mathcal L}_V =
F_\sigma^2 \, \mbox{tr} \left[
  \left\{ \partial^\mu\phi_\sigma
  {}- V^\mu + {\cal V}^\mu \right\}^2
\right] + \cdots\ ,
\end{equation}
where
\begin{equation}
F_\sigma^2 = a F^2 \ .
\end{equation}
Then, the would-be NG boson absorbed into the longitudinal
component of the $V_\mu$ is identified as
\begin{equation}
\sigma = F_\sigma \, \phi_\sigma \,.
\label{phi_sigma}
\end{equation}
The remaining three terms, 
$b{\mathcal L}_A + c {\mathcal L}_M + d {\mathcal L}_\pi$,
are expressed as
\begin{eqnarray}
&&
  b {\mathcal L}_A + c {\mathcal L}_M + d {\mathcal L}_\pi 
\nonumber\\
&&
=
(b + d)F^2\,
\mbox{tr}\bigl[ (\partial_\mu\phi_\perp)^2 \bigr]
\nonumber\\
&&\quad
{}+ (c + d)F^2\,\mbox{tr}\bigl[ ( \partial_\mu\phi_p )^2 \bigr]
{}+ (b + c)F^2\,\mbox{tr}\bigl[ ( A_\mu )^2 \bigr]
\nonumber\\
&&\quad
{}- 2 b F^2\,\mbox{tr}\bigl[ A_\mu \partial^\mu\phi_\perp \bigr]
{}+ 2c F^2\,\mbox{tr}\bigl[ A_\mu \partial^\mu\phi_p \bigr]
\nonumber\\
&&\quad
{}+ 2 d F^2\,\mbox{tr}\bigl[ \partial_\mu\phi_\perp \,
  \partial^\mu\phi_p \bigr]
 + \cdots 
\,.
\end{eqnarray}
This can be further reduced into
\begin{eqnarray}
&&
\bigl( b {\mathcal L}_A + c {\mathcal L}_M + d {\mathcal L}_\pi
\bigr)_{\rm kin}
\nonumber\\
&&
=
(b + c)F^2\,\mbox{tr}\bigl[ (A_\mu + \partial\phi_q )^2 \bigr]
\nonumber\\
&&\qquad\quad
{}+ \Bigl( d + \frac{bc}{b + c} \Bigr)F^2\,
  \mbox{tr}\bigl[ ( \partial\phi_\pi )^2 \bigr]\,,
\label{lag p^2 2}
\end{eqnarray}
where we define $\phi_\pi$ and $\phi_q$ as
\begin{eqnarray}
&&
\phi_\pi = \phi_\perp + \phi_p\,,
\nonumber\\
&&
\phi_q = \frac{1}{(b + c)}\Bigl[ c\,\phi_p
{}- b \,\phi_\perp \Bigr]\,.
\label{phi_pi q}
\end{eqnarray}
The properly normalized fields are given by
\begin{equation}
\pi = F_\pi \, \phi_\pi \ , \quad
q = F_q \, \phi_q \ ,
\end{equation}
where $F_\pi$ is the $\pi$ decay constant and
$F_q$ is the decay constant of the would-be NG boson $q$.
They are defined as
\begin{eqnarray}
&&
F_\pi^2 = (d + c\,\zeta)F^2\,,
\nonumber\\
&&
F_q^2 = (b + c)F^2\,,
\label{decay const}
\end{eqnarray}
where
\begin{equation}
\zeta = \frac{b}{b + c}\,.
\label{def zeta}
\end{equation}


Let us 
introduce the $\rho$ ($\rho$ meson and its flavor partners) and
$A_1$ ($a_1$ meson and its flavor partners) as
\begin{equation}
V^\mu = g \rho^\mu\,,
\qquad
A^\mu = g A_1^\mu\,.
\end{equation}
Then, 
expanding the Lagrangian~(\ref{lag p^2}) in terms of 
the $\pi, V_\mu$ and $A_\mu$ fields taking the unitary gauge 
$\phi_q = \phi_\sigma = 0$~\footnote{%
 When the gauge is fixed by
 taking $\xi_M = 1$ and $\xi_R = \xi_L^\dag = e^{i\pi/F_\pi}$,
 the $A_1$-$\pi$ mixing is dissolved afterwards, as shown 
 in Refs.~\cite{BKY:NPB,BKY:PRep,BFY:GHLS}.
 In this paper, on the other hand, we introduced $\pi$ 
 field to eliminate the $A_1$-$\pi$ mixing, and fixed the gauge
 to the unitary gauge by taking
 \begin{eqnarray}
 &&
 \xi_M = \exp\left[{2i\,\zeta\,\phi_\pi}\right]\,,
 \nonumber\\
 &&
 \xi_{R} = \xi_{L} = \exp\left[{i (1-\zeta)\phi_\pi }\right]\,.
 \nonumber
 \end{eqnarray}
 As emphasized in Refs.~\cite{BKY:NPB,BKY:PRep}
 the above parameterization is converted into $\xi_M=1$ and
 $\xi_R = \xi_L^\dagger = e^{i\pi/F_\pi}$ by the 
 ``gauge transformation'':
 \begin{displaymath}
   g_R = g_L^\dag = \exp\left[{i\,\zeta\,\phi_\pi}\right]\,,
 \end{displaymath}
 as
 \begin{eqnarray}
 &&
 \xi_M^\prime = g_L\, \xi_M \,g_R^\dag = 1 \,,
 \nonumber\\
 &&
 \xi_{R}^\prime = g_R\, \xi_{R} = e^{i \phi_\pi }\,,
 \nonumber\\
 &&
 \xi_{L}^\prime = g_L\, \xi_{L} = e^{-i \phi_\pi }\,.
 \nonumber
 \end{eqnarray}
},
we find the following expressions for
the masses of vector and axial-vector mesons $M_{\rho,A_1}$,
the $\rho$-$\gamma$ mixing strength $g_{\rho}$~\footnote{%
 The photon field $A_\mu$ for $N_f = 3$ is 
 embedded into ${\cal V}_\mu$ as
 \begin{displaymath}
   {\cal V}_\mu = e A_\mu Q,\quad
    Q = \left(\begin{array}{ccc}
        2/3 &      &  \\
            & -1/3 &  \\
            &      & -1/3
        \end{array}\right)\,,
 \end{displaymath}
 with $e$ being the coupling of the external gauge bosons.
}
and
strength of the coupling of the $A_1$ meson to the axial-vector
current $g_{A_1}$:
\begin{eqnarray}
&&
M_\rho = g F_\sigma \,,
\qquad
M_{A_1} = g F_q \,,
\nonumber\\
&&
g_\rho = g F_\sigma^2 \,,
\qquad\quad
g_{A_1} = g bF^2 \,.
\label{g_rhopipi tree}
\end{eqnarray}


\section{Chiral Perturbation Theory with the GHLS}
\label{sec:CHPT}

In this section, we construct the chiral perturbation theory 
(ChPT) based on the generalized hidden local symmetry (GHLS).

\subsection{General Concept}
\label{ssec:GC}

In the HLS theory, thanks to the gauge invariance,
it is possible to perform the derivative
expansion systematically.
In this ChPT with HLS~(See, for a review, Ref.~\cite{HY:PRep}),
the vector meson mass is considered as small
compared with the chiral symmetry breaking scale 
$\Lambda_\chi$, by assigning ${\cal O}(p)$ to 
the HLS gauge coupling~\cite{Georgi,Tanabashi}: 
\begin{equation}
 g \sim {\cal O}(p)\,.
\end{equation}
We adopt the same order assignment for both $\rho$ and $A_1$
mesons in the GHLS, i.e., $m_\rho \sim m_{A_1} \sim {\cal O}(p)$.
Using the above counting rule, we can systematically incorporate
the quantum corrections to several physical quantities.

In the following, we examine the smallness of our expansion
parameter $m_{\rho,A_1}/\Lambda_\chi$.
Similarly to the smallness of $m_{\rho}/\Lambda_\chi$
discussed in Ref.~\cite{HY:PRep},
the smallness of the expansion parameters
$m_{\rho,A_1}/\Lambda_\chi \ll 1$ can be
justified in a large number of colors $N_c$ of QCD as follows:
In the large $N_c$ limit, the pion decay constant $F_\pi$ 
scales as $\sqrt{N_c}$
which implies that $\Lambda_\chi$ scales as
$\Lambda_\chi \sim 4 \pi F_\pi \sim \sqrt{N_c}$.
On the other hand,
the masses of vector and axial-vector mesons, $m_{\rho,A_1}$,
do not scale with $N_c$.
So the ratios $m_{\rho,A_1}^2/F_\pi^2$ scales as $1/N_c$,
and becomes small in the large $N_c$ QCD:
\begin{equation}
\frac{m_{\rho,A_1}^2}{\Lambda_\chi^2}
= \frac{m_{\rho,A_1}^2}{(4 \pi F_\pi)^2}
\ \sim\ \frac{1}{N_c} \ \ll\ 1 .
\end{equation}
Thus
we can perform the derivative expansion 
in the large $N_c$ limit, and extrapolate the results to the 
real-life QCD
with $N_c = 3$.


\subsection{One-loop Calculations}
\label{ssec:1-loop}

Let us
calculate the quantum corrections from
the $\pi$, $\rho$ and $A_1$ meson loops to five leading-order
parameters of the GHLS Lagrangian.
We make the quantization using the background field gauge
in 't\,Hooft-Feynman gauge, which is 
summarized in Appendix~\ref{app:BFG}.

We would like to stress that
it is important to include the quadratic divergences to 
obtain the RGEs in the Wilsonian sense.
In this paper, following Refs.~\cite{HY:conformal,HY:WM,HY:PRep},
we adopt the dimensional regularization and identify 
the quadratic divergences with the presence of poles of 
ultraviolet origin at $n=2$~\cite{Veltman}.
This can be done by the following replacement in the Feynman
integrals:
\begin{eqnarray}
&&
\int \frac{d^n k}{i (2\pi)^n} \frac{1}{-k^2} 
\rightarrow 
\frac{\Lambda^2} {(4\pi)^2}\,,
\nonumber\\
&&
\int \frac{d^n k}{i (2\pi)^n} 
\frac{k_\mu k_\nu}{\left[-k^2\right]^2} 
\rightarrow 
- \frac{\Lambda^2} {2(4\pi)^2} g_{\mu\nu}\,.
\label{quad}
\end{eqnarray}
On the other hand, 
the logarithmic divergence is identified with the pole at $n=4$:
\begin{equation}
\frac{1}{\bar{\epsilon}} + 1 
\rightarrow
\ln \Lambda^2\,,
\label{log}
\end{equation}
where
\begin{equation}
\frac{1}{\bar{\epsilon}} \equiv
\frac{2}{4 - n } - \gamma_E + \ln (4\pi)\,,
\end{equation}
with $\gamma_E$ being the Euler constant.

In the following, we consider the
two-point functions of
$\bar{V}^\mu$-$\bar{V}^\nu$,
$\bar{A}^\mu$-$\bar{A}^\nu$,
$\bar{\cal A}_\perp^\mu$-$\bar{A}^\nu$ and
$\bar{\cal A}_{M}^\mu$-$\bar{\cal A}_\perp^\nu$
(see Appendix~\ref{app:BFG} for definitions of the 
background fields),
which we express 
as $\Pi_{\bar{V}\bar{V}}^{\mu\nu}$, 
$\Pi_{\bar{A}\bar{A}}^{\mu\nu}$,
$\Pi_{\bar{\cal A}_\perp\bar{A}}^{\mu\nu}$,
$\Pi_{\bar{\cal A}_{M}\bar{\cal A}_\perp}^{\mu\nu}$,
respectively.
We divide each of these two-point functions into two parts as
\begin{equation}
\Pi^{\mu\nu}(p) = \Pi^S(p^2) g^{\mu\nu} 
{}+ \Pi^{LT}(p^2)(p^2g^{\mu\nu} - p^\mu p^\nu)\,.
\end{equation}
At the bare level, the relevant parts are expressed as
\begin{eqnarray}
&&
\Pi_{\bar{V}\bar{V}}^{{\rm (bare)}S}
= a_{\rm bare}F^2\,,
\nonumber\\
&&
\Pi_{\bar{V}\bar{V}}^{{\rm (bare)}LT}
= - \frac{1}{g_{\rm bare}^2} + 2 z^{LR}_{\rm bare}\,,
\nonumber\\
&&
\Pi_{\bar{A}\bar{A}}^{{\rm (bare)}S}
= (b_{\rm bare} + c_{\rm bare})F^2\,,
\nonumber\\
&&
\Pi_{\bar{A}\bar{A}}^{{\rm (bare)}LT}
= - \frac{1}{g_{\rm bare}^2} - 2 z^{LR}_{\rm bare}\,,
\nonumber\\
&&
\Pi_{\bar{\cal A}_\perp\bar{A}}^{{\rm (bare)}S}
= - b_{\rm bare}F^2\,,
\nonumber\\
&&
\Pi_{\bar{\cal A}_{M}\bar{\cal A}_\perp}^{{\rm (bare)}S}
= d_{\rm bare}F^2\,,
\label{bare}
\end{eqnarray}
where $z_{\rm bare}^{LR}$ is the coefficient of ${\cal O}(p^4)$
terms which proportional to 
$\mbox{tr}\left[ L_{\mu\nu}\xi_M R^{\mu\nu}\xi_M^\dagger \right]$
~\footnote{%
 Some of the possible ${\mathcal O}(p^4)$  terms 
 contributing to three- and four-point functions 
 are listed in Refs.~\cite{BKY:PRep,Wu:GHLS}.
 A complete list of the anomalous terms at ${\mathcal O}(p^4)$
 is given in Ref.~\cite{Kaiser:1990yf}.
}.

In Appendix~\ref{app:QC},
we list the diagrams contributing to $\Pi^{\mu\nu}$ 
at one-loop level and the quantum corrections from those diagrams
(see Figs.~\ref{fig:VV}-\ref{fig:Mperpperp} and
Eqs.~(\ref{eq:appB:1})-(\ref{eq:appB:4})).
{}From Eq.~(\ref{bare}), we find that
the divergences proportional to $g^{\mu\nu}$ in the
two-point functions are renormalized by 
$a_{\rm bare}, b_{\rm bare}, c_{\rm bare}$ and $d_{\rm bare}$
and those proportional to $(p^2 g^{\mu\nu} - p^\mu p^\nu)$
are renormalized by $g_{\rm bare}$ and $z_{\rm bare}^{LR}$.
Thus we require the following renormalization conditions:
\begin{eqnarray}
&&
a_{\rm bare}F^2 + \Pi_{\bar{V}\bar{V}}^S \big|_{\rm div}
= (\mbox{finite})\,,
\nonumber\\
&&
- b_{\rm bare}F^2 + \Pi_{\bar{\cal A}_\perp\bar{A}}^S \big|_{\rm div}
= (\mbox{finite})\,,
\nonumber\\
&&
c_{\rm bare}F^2 + \Pi_{\bar{A}\bar{A}}^S \big|_{\rm div}
{}+ \Pi_{\bar{\cal A}_\perp\bar{A}}^S \big|_{\rm div}
= (\mbox{finite})\,,
\nonumber\\
&&
d_{\rm bare}F^2 
{}+ \Pi_{\bar{\cal A}_{M}\bar{\cal A}_\perp}^S \big|_{\rm div}
= (\mbox{finite})\,,
\nonumber\\
&&
- \frac{1}{g_{\rm bare}^2} 
{}+ \frac{1}{2}\Bigl[ \Pi_{\bar{V}\bar{V}}^{LT} \big|_{\rm div}
{}+ \Pi_{\bar{A}\bar{A}}^{LT} \big|_{\rm div} \Bigr]
= (\mbox{finite})\,.
\nonumber\\
\label{renormalization}
\end{eqnarray}
{}From the above renormalization conditions, we obtain
the renormalization group equations (RGEs) for
the parameters $a$, $b$, $c$, $d$ and the GHLS gauge coupling
$g$, which are listed in Eqs.~(\ref{eq:appB:5})-(\ref{eq:appB:9}).


\section{Weinberg's Sum Rules}
\label{sec:WSR}

Let us start with 
the axial-vector and vector current correlators defined by
\begin{eqnarray}
&&
G_A(Q^2)(q^\mu q^\nu - q^2 g^{\mu\nu}) \delta_{ab}
\nonumber\\
&&\qquad\quad
=
\int d^4x\,e^{iqx}
\left\langle 0 \vert \, T\,
  J_{5a}^\mu(x) J_{5b}^\nu(0)
\vert 0 \right\rangle\,,
\nonumber\\
&&
G_V(Q^2)(q^\mu q^\nu - q^2g^{\mu\nu}) \delta_{ab}
\nonumber\\
&&\qquad\quad
=
\int d^4x\,e^{iqx}
\left\langle 0 \vert \, T\,
  J_a^\mu(x) J_b^\nu(0)
\vert 0 \right\rangle\,,
\end{eqnarray}
where $Q^2 = - q^2$ is the space-like momentum,
$J_{5a}^\mu$ and $J_a^\mu$ are the axial-vector and 
vector currents and $(a,b)=1,\ldots,N_f^2-1$ denotes the flavor
index. 
At the leading order of the GHLS
the current correlators $G_{A,V}$ 
are expressed as
\begin{eqnarray}
&&
G_A(Q^2)
= \frac{F_\pi^2}{Q^2} + \frac{F_{A_1}^2}{M_{A_1}^2+Q^2}\,,
\nonumber\\
&&
G_V(Q^2)
= \frac{F_\rho^2}{M_\rho^2 + Q^2}\,,
\label{pole}
\end{eqnarray}
where the $A_1$ and $\rho$ decay constants are defined by
\begin{eqnarray}
&&
F_{A_1}^2 = \Bigl( \frac{g_{A_1}}{M_{A_1}} \Bigr)^2 
= \frac{b^2}{b+c}F^2\,,
\nonumber\\
&&
F_\rho^2 = \Bigl( \frac{g_\rho}{M_\rho} \Bigr)^2 
= a F^2\,.
\end{eqnarray}

The same correlators are evaluated by the OPE as~\cite{SVZ}
\begin{eqnarray}
&&
G_A^{\rm(OPE)}(Q^2) = \frac{1}{8\pi^2}
\Biggl[
  - \left( 1 + \frac{\alpha_s}{\pi} \right) \ln \frac{Q^2}{\mu^2}
\nonumber\\
&& \quad
  + \frac{\pi^2}{3} 
    \frac{
      \left\langle 
        \frac{\alpha_s}{\pi} G_{\mu\nu} G^{\mu\nu}
      \right\rangle
    }{ Q^4 }
  + \frac{\pi^3}{3} \frac{1408}{27}
    \frac{\alpha_s \left\langle \bar{q} q \right\rangle^2}{Q^6}
\Biggr]
\ ,
\nonumber\\
&&
G_V^{\rm(OPE)}(Q^2) = \frac{1}{8\pi^2}
\Biggl[
  - \left( 1 + \frac{\alpha_s}{\pi} \right) \ln \frac{Q^2}{\mu^2}
\nonumber\\
&& \quad
  + \frac{\pi^2}{3} 
    \frac{
      \left\langle 
        \frac{\alpha_s}{\pi} G_{\mu\nu} G^{\mu\nu}
      \right\rangle
    }{ Q^4 }
  - \frac{\pi^3}{3} \frac{896}{27}
    \frac{\alpha_s \left\langle \bar{q} q \right\rangle^2}{Q^6}
\Biggr]
\ ,
\label{Pi A V OPE}
\end{eqnarray}
where $\mu$ is the renormalization scale of QCD.
An important result obtained from the above forms is that
the difference between two correlators scales as $1/Q^6$:
\begin{equation}
G_A^{\rm(OPE)}(Q^2) - G_V^{\rm(OPE)}(Q^2) 
= \frac{32\pi}{9} \frac{\alpha_s \,\langle \bar{q}q\rangle^2}{Q^6}
\ .
\label{A-V:OPE}
\end{equation}

Since the above forms of the correlators in the OPE are valid
in the high energy region, we consider the difference of the
correlators in the GHLS in the energy region higher than the
$A_1$ meson mass, i.e., $Q^2 \gg M_A^2$.
In the high energy region, 
two correlators in the GHLS given in Eq.~(\ref{pole})
are expanded as
\begin{eqnarray}
&&
G_A(Q^2) = \frac{F_\pi^2 + F_{A_1}^2}{Q^2}
{}- \frac{F_{A_1}^2 M_{A_1}^2}{Q^4}
{}+ \frac{F_{A_1}^2 M_{A_1}^4}{Q^6}\,,
\nonumber\\
&&
G_A(Q^2) = \frac{F_\rho^2}{Q^2}
{}- \frac{F_\rho^2 M_\rho^2}{Q^4}
{}+ \frac{F_\rho^2 M_\rho^4}{Q^6}\,.
\end{eqnarray}
{}From the above expressions, the difference of two correlators
is given by
\begin{eqnarray}
\lefteqn{
G_A(Q^2) - G_V(Q^2) 
=
\frac{ F_\pi^2 + F_{A_1}^2 - F_\rho^2 }{Q^2}
}
\nonumber\\
&& {}
+ \frac{ F_{A_1}^2 M_{A_1}^2 - F_\rho^2 M_\rho^2 }{Q^4}
+ \frac{F_{A_1}^2 M_{A_1}^4 - F_\rho^2 M_\rho^4}{Q^6}
\,.
\nonumber\\
\label{A-V:GHLS}
\end{eqnarray}

We require that the high energy behavior of the difference between
two correlators in the GHLS agrees with that in the OPE:
$G_A(Q^2) - G_V(Q^2)$ in the GHLS scales as $1/Q^6$.
This requirement can be satisfied only if the following relations
are satisfied:
\begin{eqnarray}
&&
F_\pi^2 + F_{A_1}^2 = F_\rho^2\,,
\nonumber\\
&&
F_{A_1}^2 M_{A_1}^2 = F_\rho^2 M_\rho^2\,,
\label{WSR}
\end{eqnarray}
which are nothing but the pole saturated forms of the 
Weinberg first and second sum rules~\cite{Weinberg}.
In terms of the parameters of the GHLS Lagrangian,
the above relations can be satisfied
if we take 
\begin{equation}
a = b \,, \quad d = 0 \,.
\label{tsl}
\end{equation}

Now, let us study whether the above relations in Eq.~(\ref{tsl})
are stable against the quantum corrections.
Taking $a=b$ and $d=0$ in the RGEs for $a$, $b$ and $d$ shown in
Eqs.~(\ref{eq:appB:6}), (\ref{eq:appB:7}) and (\ref{eq:appB:8}),
we obtain
\begin{eqnarray}
\mu\frac{d (a F^2)}{d \mu}
&=& \frac{N_f}{(4\pi)^2}\Bigl[ \mu^2 + 3 a g^2 F^2 \Bigr]\,,
\label{RGE a}
\\
\mu\frac{d (b F^2)}{d \mu}
&=& \frac{N_f}{(4\pi)^2}\Bigl[ \mu^2 + 3 a g^2 F^2 \Bigr]\,,
\label{RGE b}
\\
\mu\frac{d (d F^2)}{d \mu}
&=& 0\,.
\label{RGE d}
\end{eqnarray}
The first two RGEs lead to
\begin{equation}
\mu\frac{d (a - b )}{d \mu}
= 0\,.
\label{RGE a-b}
\end{equation}
The RGEs in Eqs.~(\ref{RGE d}) and (\ref{RGE a-b}) imply that 
the parameter relations $a = b$ and $d=0$ are stable
against the renormalization group evolution,
i.e., {\it the non-renormalization of the Weinberg sum rules
expressed in terms of the leading order parameters
in the GHLS}.

At the last of this section,
we look into a set of the parameter relations
in Eq.~(\ref{tsl})
to the symmetry structure of the GHLS theory.
When we take $a=b$ and $d=0$,
the
GHLS Lagrangian given in Eq.~(\ref{lag a-d}) is rewritten as
\begin{align}
&
a{\cal L}_V + b{\cal L}_A + c{\cal L}_M
{}+ d{\cal L}_\pi 
\nonumber\\
&
= - 8 a F^2 \,\mbox{tr}\Bigl[ \bigl( D_\mu \xi_R \bigr)^2
  {}+ \bigl( D_\mu \xi_L \bigr)^2 \Bigr]
\nonumber\\
&\qquad
 - 4 c F^2 \, \mbox{tr}\left[ \bigl( D_\mu \xi_M \bigr)^2 \right]\,.
\end{align}
When we further switch off the gauge coupling,
the symmetry of the Lagrangian becomes enhanced as
$G_{\rm global}\times [G_{\rm global}]^2 = [G_{\rm global}]^3
= [ SU(N_f)_L\times SU(N_f)_R ]^3$.
This implies that three variables $\xi_L$, $\xi_R$ and $\xi_M$
couple to each other only through the GHLS gauge bosons
$V_\mu$ and $A_\mu$, when the gauge coupling is switched on.
This structure is generally refereed as the ``theory space 
locality''~\cite{deconstruction,HTY,SS,Piai}.
{}From the above consideration, we see that,
in the GHLS, the requirement of the Weinberg sum rules
automatically leads to the 
``theory space locality''~\cite{CKT}.
In general cases, the ``theory space locality'' is
satisfied only at tree level, since the enhanced symmetry
is broken when the gauge coupling is switched on.
However, our result of the stability of the relations
$a=b$ and $d=0$ implies that 
{\it the  
``theory space locality'' in the leading order Lagrangian
is stable against the quantum
correction} at least at one-loop level.


\section{Chiral Symmetry Restoration}
\label{sec:PS}

In this section, we 
study 
the chiral phase transition 
keeping the first and second Weinberg sum rules in the
GHLS.

Equations~(\ref{A-V:OPE}) and (\ref{A-V:GHLS}) with
the Weinberg sum rules (\ref{WSR}) 
give the following
matching condition in the high-energy region:
\begin{eqnarray}
&&
F_{A_1}^2 M_{A_1}^4 - F_\rho^2 M_\rho^4
\nonumber\\
&& \quad
=
F_\rho^2 M_\rho^2 \left( M_{A_1}^2 - M_\rho^2 \right) 
=
\frac{32\pi\alpha_s}{9} \langle \bar{q}q \rangle^2\,,
\label{match rest}
\end{eqnarray}
which is a measure of the spontaneous chiral symmetry breaking.
In terms of the parameters of the GHLS Lagrangian
this is expressed as
\begin{equation}
a^2 \cdot c \cdot g^4 \propto \langle \bar{q}q \rangle^2\,.
\label{matching}
\end{equation}
When the chiral restoration point is approached,
the quark condensate approaches zero:
\begin{equation}
\langle \bar{q} q \rangle \ \rightarrow\  0 \,,
\end{equation}
This implies that the condition
\begin{equation}
a^2 \cdot c \cdot g^4 \ \rightarrow\  0 
\end{equation}
is satisfied when the chiral symmetry is restored.
{}From this condition we see that
at least one parameter among $a, c$ and $g$ must go to zero
at the chiral symmetry restoration point.

Let us first consider the possibility that the parameter $a$
goes to zero at a high energy scale, say $\Lambda$:
$a(\Lambda) \rightarrow 0$.
The RGE for $a$ given in Eq.~(\ref{RGE a}) implies that
$a=0$ is not a fixed point, and thus
one cannot achieve the equality of the axial-vector and vector
current correlator in the energy region below $\Lambda$
which is required by the chiral symmetry restoration.
To make the matters worse,
the RGE for $a$ leads to
$a(\mu) < 0$ for $\mu < \Lambda$, and $M_\rho^2 < 0$,
which is of course unacceptable.
{}From these, we cannot take $a \rightarrow 0$.

We next consider the possibility of $c(\Lambda) \rightarrow 0$.
{}From Eq.~(\ref{eq:appB:7}),
the RGE for $c$ in the case of $a=b$ and $d=0$ is obtained as
\begin{equation}
\mu\frac{d (c F^2)}{d \mu}
= \frac{N_f}{(4\pi)^2}\Bigl[ 2\mu^2 + 6c g^2 F^2 \Bigr]\,.
\label{RGE c}
\end{equation}
We can easily see that $c=0$ is not a fixed point,
which implies that
the equality of two current correlators cannot be satisfied
in the energy region below $\Lambda$ even if we equate them
at $\Lambda$.
Furthermore,
negative $c$ leads to
$M_\rho^2 / M_{A_1}^2 = a/(a+c) > 1$,
which is unacceptable, either.
Thus $c \rightarrow 0 $ cannot be achieved at the
restoration point.

Finally, we study the possibility of $g \rightarrow 0$.
{}From Eq.~(\ref{eq:appB:9}), 
the RGE for $g$ with $a=b$ and $d=0$ is reduced to
\begin{equation}
\mu\frac{d g^2}{d \mu}
= - \frac{N_f}{(4\pi)^2}\frac{43}{3}g^4\,,
\label{RGE g}
\end{equation}
which certainly has the fixed point at $g = g^\ast = 0$.
Then, the symmetry restoration in the GHLS 
can be realized only if
the following condition is met:
\begin{equation}
 g \ \rightarrow\ g^\ast = 0 \,.
\label{cond}
\end{equation}
This condition implies the massless $\rho$ and $A_1$ mesons,
since both masses are proportional to the gauge coupling
$g$.~\footnote{%
 This symmetry restoration is similar to the vector 
 manifestation (VM)~\cite{HY:VM,HY:PRep}, in which the
 massless $\rho$ becomes the chiral partner of the pion.
 As is stressed for the VM in Ref.~\cite{HY:PRep},
 the symmetry restoration here should also be considered only as a
 limit with bare parameters approaching the fixed point:
 An enhancement of the global symmetry occurs when we take
 $g=0$ from the beginning. While for non-zero gauge coupling,
 even if it is very tiny, the global symmetry in the GHLS is
 only the chiral symmetry consistently with QCD.
}
Thus we conclude that,
{\it when we require the first and second Weinberg sum rules to be
satisfied,
the chiral symmetry restoration in the GHLS 
required through the matching to QCD 
can be realized with masses of $\rho$ and $A_1$ mesons vanishing
at the restoration point}:
\begin{equation}
M_\rho \ \rightarrow\ 0 \,, \quad
M_{A_1}\ \rightarrow\ 0 \,.
\end{equation}

We next consider the fate of two parameters $a$ ($=b$) and $c$.
As we can see easily from Eq.~(\ref{match rest}) or 
Eq.~(\ref{matching}),
matching of the GHLS to QCD does not provide any conditions
for $a$ and $c$ other than $g\rightarrow 0$ at the restoration 
point.
For $a=b$ and $d=0$
the definitions of the parameter $\zeta$ and the pion
decay constant given in
Eqs.~(\ref{decay const}) and (\ref{def zeta}) 
are rewritten as
\begin{eqnarray}
F_\pi^2 &=& \frac{a c}{a+c} F^2 \,,
\label{decay constant 2}
\\
\zeta &=& \frac{a}{a+c} = \frac{M_\rho^2}{M_{A_1}^2}\,,
\label{mass ratio}
\end{eqnarray}
{}From this, we see that
the parameter $\zeta$ plays an important role,
which 
controls the fate of the ratio of $\rho$ and $A_1$ meson masses
at the symmetry restoration.
Below,
we shall investigate the phase structure of the GHLS
to see how the mass ratio $\zeta$ is determined at the
symmetry restoration point and
characterizes the possible patterns of chiral symmetry restoration
governed by several fixed points.

To study the phase structure of the GHLS through the RGEs for
$a$, $c$ and $g$, it is convenient to
introduce the following dimensionless parameters
associated with $a, c$ and $g$:
\begin{eqnarray}
&&
X(\mu) = \frac{N_f}{2(4\pi)^2}\frac{\mu^2}{a(\mu)F^2}\,,
\nonumber\\
&&
Y(\mu) = \frac{N_f}{2(4\pi)^2}\frac{\mu^2}{c(\mu)F^2}\,,
\nonumber\\
&&
G(\mu) = \frac{N_f}{2(4\pi)^2}g^2(\mu)\,.
\end{eqnarray}
In terms of $X$ and $Y$, the order parameter $F_\pi$ and 
the mass ratio $\zeta$ are expressed as
\begin{eqnarray}
&&
W(\mu) = \frac{N_f}{2(4\pi)^2}\frac{\mu^2}{F_\pi^2(\mu)}
 = X(\mu) + Y(\mu)\,,
\label{def W}
\\
&&
\zeta(\mu) = \frac{Y(\mu)}{X(\mu) + Y(\mu)}\,. 
\end{eqnarray}
The RGEs shown in
Eqs.~(\ref{RGE a}), (\ref{RGE c}) and (\ref{RGE g}) are rewritten 
as
\begin{eqnarray}
&&
\mu\frac{d X}{d \mu}
= 2X(1 - X - 3G)\,,
\nonumber\\
&&
\mu\frac{d Y}{d\mu}
= 2Y(1 - 2Y - 6G)\,,
\nonumber\\
&&
\mu\frac{d G}{d\mu}
= {}- \frac{86}{3}G^2\,.
\end{eqnarray}
{}From these RGEs
we find that three non-trivial fixed points and one trivial 
fixed point.
The trivial fixed point is given by
\begin{equation}
(X^\ast,Y^\ast,G^\ast) = (0,0,0)\,,
\end{equation}
while non-trivial ones are
\begin{eqnarray}
&&\mbox{A :}\quad
(X^\ast,Y^\ast,G^\ast) = (1,0,0)\,,
\nonumber\\
&&\mbox{B :}\quad
(X^\ast,Y^\ast,G^\ast) = (0,1/2,0)\,,
\nonumber\\
&&\mbox{C :}\quad
(X^\ast,Y^\ast,G^\ast) = (1,1/2,0)\,.
\label{FP-XY}
\end{eqnarray}

\begin{figure}
 \begin{center}
  \includegraphics[width = 8cm]{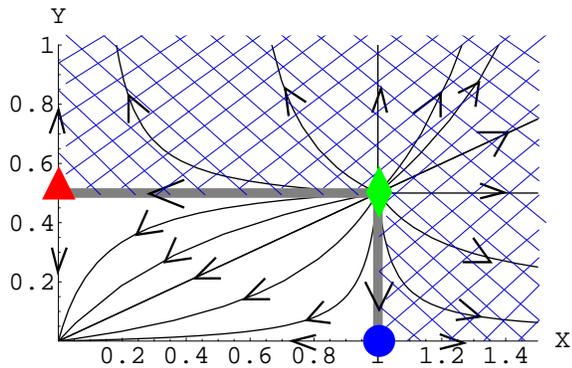}
 \end{center}
 \caption{Phase diagram on $G=0$ plane. Arrows on the flows are
  written from the ultraviolet to the infrared. Gray lines divide
  the broken phase (inside) and the symmetric phase 
  (outside; cross-hatched area).
  Points denoted by $\blacktriangle$, $\bullet$ and $\blacklozenge$ 
  express the fixed point $(X,Y)=(0,1/2)$, $(1,0)$ and $(1,1/2)$
  respectively.}
 \label{fig:XY}
\end{figure}

As we concluded above, the symmetry restoration can be
realized only if we have $G\rightarrow0$ at the restoration point.
Since $G = 0$ is the only fixed point of the RGE for $G$,
we concentrate on the case with $G=0$.
In such a case, the RGE flows are confined on the $X$-$Y$ plane.
Furthermore,
since both $\rho$ and $A_1$ mesons are massless,
we can use the RGEs for $X$ and $Y$ all the way down to the
low-energy limit, $\mu = 0$.
Then,
the phase of the GHLS is determined by the on-shell pion 
decay constant $F_\pi(\mu = 0)$,
or equivalently $W$ defined in Eq.~(\ref{def W}), 
as
\begin{eqnarray}
&&
W(\mu=0)=0 \qquad \mbox{broken phase}
\nonumber\\
&&
W(\mu=0)\neq 0 \qquad \mbox{symmetric phase}
\end{eqnarray}
We show the flow diagram in $X$-$Y$ plane in Fig.~\ref{fig:XY}.
The phase boundary is specified by $F_\pi(0)=0$ which is realized
at each fixed point listed in Eq.~(\ref{FP-XY}).
The fixed point A implies $a(0)=0$ and $c(0)\neq 0$,
B entails $a(0)\neq 0$ and $c(0)=0$,
and C gives us $a(0)=c(0)=0$.
We note that $a(0)=0$ and/or $c(0)=0$ are realized
due to the quadratic running of the RGEs although the bare
parameters $a_{\rm bare}$ and $c_{\rm bare}$ are non-zero
even at the restoration point.

\begin{figure}
 \begin{center}
  \includegraphics[width = 8cm]{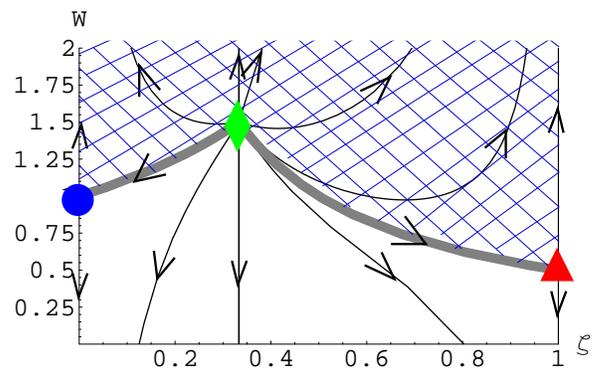}
 \end{center}
 \caption{Phase diagram on $\zeta$-$W$ plane. Arrows on the flows are
  written from the ultraviolet to the infrared. Gray lines divide
  the broken phase (lower side) and the symmetric phase 
  (upper side; cross-hatched area).
  Points denoted by $\blacktriangle$, $\bullet$ and $\blacklozenge$ 
  express the fixed point $(\zeta,W)=(1,1/2)$, $(0,1)$ and $(1/3,3/2)$
  respectively.}
 \label{fig:zW}
\end{figure}

In order to clarify the implication of each fixed point,
we map the phase diagram in the $X$-$Y$ plane onto 
the $\zeta$-$W$ plane, which is shown in Fig.~\ref{fig:zW}.
Three fixed points (\ref{FP-XY}) are turned into
\begin{eqnarray}
&&\mbox{A :}\quad
(\zeta^\ast,W^\ast)=(1,1/2)\,,
\nonumber\\
&&\mbox{B :}\quad
(\zeta^\ast,W^\ast)=(0,1)\,,
\nonumber\\
&&\mbox{C :}\quad
(\zeta^\ast,W^\ast)=(1/3,3/2)\,.
\label{FP-zW}
\end{eqnarray}
{}From this we can distinguish three patterns of the
chiral symmetry restoration characterized by three fixed points
by the values of the ratio of $\rho$ and $A_1$ meson masses
expressed by $\zeta$ as in Eq.~(\ref{mass ratio}) as follows:
At the fixed point A,
$\zeta$ goes to 1, which implies that the $\rho$ meson mass
degenerates into the $A_1$ meson mass.
We shall call this restoration pattern
the Ginzburg-Landau (GL) type. 
At the fixed point B, on the other hand,
the $\rho$ meson becomes massless faster than the $A_1$ meson
since $\zeta$ goes to zero.
This can be called the vector manifestation (VM) type.
The fixed point C is the ultraviolet fixed point in any direction,
so that it is not so stable as to A and B.
Nevertheless, if it is chosen,
the mass ratio approaches to $1/3$ which we shall call the
hybrid type.

To summarize, we find that
the chiral symmetry restoration in the GHLS 
required through the matching to QCD can be realized 
only if the masses of $\rho$ and $A_1$ mesons vanish
at the restoration point:
\begin{equation}
M_\rho \ \rightarrow\ 0 \,, \quad
M_{A_1}\ \rightarrow\ 0 \,,
\end{equation}
and that
the ratio of these masses flows into one of the 
following three fixed points:
\begin{eqnarray}
&&
\mbox{GL-type} \ : \ 
  M_\rho^2/M_{A_1}^2 \ \rightarrow\ 1 \,,
\nonumber\\
&&
\mbox{VM-type} \ : \ 
  M_\rho^2/M_{A_1}^2 \ \rightarrow\ 0 \,,
\nonumber\\
&&
\mbox{Hybrid-type} \ : \ 
  M_\rho^2/M_{A_1}^2 \ \rightarrow\ 1/3 \,.
\end{eqnarray}


\section{Chiral Representation Mixing}
\label{sec:CRM}

In this section, we discuss the relation of 
three classes of the fixed point studied in previous section
to the chiral representation mixing.

In the broken phase of the chiral symmetry, 
the eigenstates of the chiral representation 
under $SU(N_f)_L \times SU(N_f)_R$ do not generally agree with
the mass eigenstates due to the existence of the Nambu-Goldstone 
bosons, i.e., there exists a representation mixing.
By extending the analysis done in 
Ref.~\cite{Gilman:1967qs,Weinberg:hw} for two-flavor QCD,
the scalar, pseudoscalar, longitudinal vector and axial-vector 
mesons belong to the following representations:
\begin{eqnarray}
 |s \rangle 
   &=& |(N_f,N_f^\ast) \oplus (N_f^\ast,N_f) \rangle\,, 
\nonumber\\
 |\pi \rangle 
   &=& |(N_f,N_f^\ast) \oplus (N_f^\ast,N_f) \rangle \sin\psi 
\nonumber\\
&&{}+
       |(1,N_f^2-1) \oplus (N_f^2-1,1) \rangle \cos\psi\,, 
\nonumber\\
 |\rho \rangle 
   &=& |(1,N_f^2-1) \oplus (N_f^2-1,1) \rangle\,, 
\nonumber\\
 |A_1 \rangle 
   &=& |(N_f,N_f^\ast) \oplus (N_f^\ast,N_f) \rangle \cos\psi 
\nonumber\\
&&{}-
       |(1,N_f^2-1) \oplus (N_f^2-1,1) \rangle \sin\psi\,,
\label{rep-mixing}
\end{eqnarray}
where $\psi$ denotes the mixing angle.
The value of $\psi$ for $N_f=2$ 
is estimated as about $\psi \simeq 45\,{}^\circ$.

It can be expected that the above representation mixing is dissolved
when the chiral symmetry is restored.
{}From Eq.~(\ref{rep-mixing}), one can easily see that 
there are two possibilities for pattern of chiral symmetry restoration.
One possible pattern is the case where $\cos\psi \to 0$ 
when we approach the critical point.
In this case, the pion belongs to 
$|(N_f,N_f^\ast) \oplus (N_f^\ast,N_f)\rangle$
and becomes the chiral partner of the scalar meson.
The longitudinal vector and axial-vector mesons are in the same
multiplet 
$|(1,N_f^2-1) \oplus (N_f^2-1,1) \rangle$.
This is the standard Ginzburg-Landau (GL) scenario of 
the chiral symmetry restoration.
Another possibility is the case where $\sin\psi \to 0$ 
when we approach the critical point.
In this case, the pion belongs to pure 
$|(1,N_f^2-1) \oplus (N_f^2-1,1)\rangle$
and its chiral partner is now the (longitudinal) vector meson.
The scalar meson joins with the longitudinal part of the
axial-vector meson 
in the same representation 
$|(N_f,N_f^\ast) \oplus (N_f^\ast,N_f) \rangle$.
This is the vector manifestation (VM) of 
chiral symmetry~\cite{HY:VM,HY:PRep}.

Now we consider how the chiral representation mixing
is expressed in the GHLS theory.
When we take $d=0$ in the Lagrangian,
there are no $\phi_\perp$-$\phi_p$ mixing terms
[see Eq.~(\ref{lag p^2 2})].
Then we take the normalizations of 
$\phi_\perp$ and $\phi_p$ fields as follows:
\begin{eqnarray}
&&
\phi_\perp = \pi_\perp / \sqrt{bF^2}\,,
\qquad
\phi_p = p / \sqrt{cF^2}\,.
\end{eqnarray}
Since $\pi_\perp$ is included in $\xi_L$ and $\xi_R$
and $p$ is in $\xi_M$,
we identify $\pi_\perp$ with the field
belonging to the chiral representation
$(1,N_f^2-1) \oplus (N_f^2-1,1)$ and $p$ with
$(N_f,N_f^\ast) \oplus (N_f^\ast,N_f)$
according to the transformation properties of 
$\xi_L$, $\xi_R$ and $\xi_M$.
Using Eqs.~(\ref{phi_pi q}) and (\ref{decay const}), 
we rewrite $\pi$ and $q$ in terms of $\pi_\perp$ and $p$ as
\begin{eqnarray}
&&
\pi = \sqrt{\zeta}\,p + \sqrt{1-\zeta}\,\pi_\perp\,,
\nonumber\\
&&
q = \sqrt{1-\zeta}\,p - \sqrt{\zeta}\,\pi_\perp\,.
\end{eqnarray}
We compare the above expression to Eq.~(\ref{rep-mixing})
and obtain that the chiral representation mixing angle
is related to the mass ratio $\zeta$ as
\begin{eqnarray}
\cos\psi = \sqrt{1-\zeta}\,,
\qquad
\sin\psi = \sqrt{\zeta}\,.
\label{angle}
\end{eqnarray}
Then, from three fixed points (\ref{FP-zW}) of the GHLS 
at the symmetry restoration point,
the fate of the chiral representation mixing is determined
as follows:
\begin{eqnarray}
&&
\mbox{GL-type} \ : \ 
  \cos \psi \ \rightarrow\ 0 \,,
\nonumber\\
&&
\mbox{VM-type} \ : \ 
  \sin \psi \ \rightarrow\ 0 \,,
\nonumber\\
&&
\mbox{Hybrid-type} \ : \ 
\nonumber\\
&& \qquad
  \sin \psi \ \rightarrow\ \sqrt{\frac{1}{3}} \,,
  \quad
  \cos \psi \ \rightarrow\ \sqrt{\frac{2}{3}} \,.
\end{eqnarray}


\section{Summary and Discussions}
\label{sec:SD}

In this paper, we developed the chiral perturbation theory
(ChPT) with the generalized hidden local symmetry (GHLS)
as an effective field theory (EFT) of QCD for pions,
vector and axial-vector mesons.
We showed that the first and second Weinberg sum rules
expressed in terms of the leading order parameters,
which is required by the equality of 
the high-energy behaviors of the current correlators of the 
GHLS to the ones in QCD,
can be satisfied by a special parameter choice, $a=b$ and $d=0$,
corresponding to the theory space locality.
Our analysis using the one-loop RGEs provides that
this parameter choice is stable against the RG evolution, 
i.e., the non-renormalization of the Weinberg sum rules
in the GHLS:
The completion of the sum rules at the bare level 
is kept even at quantum level.

With the set of parameters corresponding to the Weinberg sum rules,
we investigated the phase structure of the GHLS theory.
We found that both $\rho$ and $A_1$ meson become massless
at the chiral phase transition point, which is protected
by the fact that the GHLS gauge coupling constant $g$ goes to
zero as the only fixed point of the RGE for $g$:
\begin{equation}
M_\rho \ \rightarrow\ 0 \,, \quad
M_{A_1}\ \rightarrow\ 0 \,,
\end{equation}
At the critical point, there exist three fixed points of the RGEs
for $a$ and $c$ and each of them is associated with one of
the following patterns of the chiral symmetry restoration:
the Ginzburg-Landau (GL), the vector manifestation (VM) and
the hybrid type.
Those classes of the chiral symmetry restoration are characterized
by the mass ratio, 
or equivalently the chiral representation mixing angle,
as
\renewcommand{\arraystretch}{1.3}
\begin{eqnarray}
&&
\mbox{GL-type} \ : \ 
\left\{\begin{array}{l}
  M_\rho^2/M_{A_1}^2 \ \rightarrow\ 1 \,,
 \\
  \cos \psi \ \rightarrow\ 0 \,,
\end{array}\right.
\nonumber\\
&&
\mbox{VM-type} \ : \ 
\left\{\begin{array}{l}
  M_\rho^2/M_{A_1}^2 \ \rightarrow\ 0 \,,
 \\
  \sin \psi \ \rightarrow\ 0 \,,
\end{array}\right.
\nonumber\\
&&
\mbox{Hybrid-type} \ : \ 
\left\{\begin{array}{l}
  M_\rho^2/M_{A_1}^2 \ \rightarrow\ 1/3 \,.
 \\
  \sin \psi \ \rightarrow\ \sqrt{\frac{1}{3}} \,.
\end{array}\right.
\end{eqnarray}
\renewcommand{\arraystretch}{1}

\begin{figure}
 \begin{center}
  \includegraphics[width = 7cm]{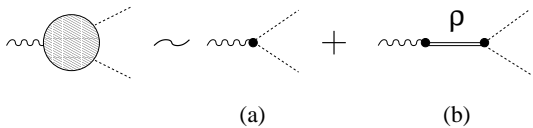}
 \end{center}
 \caption{Leading contributions to the electromagnetic form factor
  of the pion. (a) direct $\gamma\pi\pi$ and 
  (b) $\gamma\pi\pi$ mediated by $\rho$-meson exchange.
  }
 \label{fig:VD}
\end{figure}

Here we study the fate of the vector dominance (VD) 
of the electromagnetic form factor of the pion~\cite{Sakurai}
at the chiral symmetry restoration.
The direct photon-$\pi$-$\pi$ coupling $g_{\gamma\pi\pi}$ is
given by
\begin{equation}
g_{\gamma\pi\pi}
= 1 - \frac{1}{2}\frac{F_\sigma^2}{F_\pi^2} (1-\zeta^2)\,.
\end{equation}
When $a=b$ and $d=0$ are taken, this becomes
\begin{equation}
g_{\gamma\pi\pi}= (1-\zeta)/2\,. 
\end{equation}
We show the leading contributions to the pion form factor 
in Fig.~\ref{fig:VD}.
The VD is characterized by the direct $\gamma\pi\pi$ being zero.
Three classes of the chiral symmetry restoration
give us the following results on the VD:
\begin{eqnarray}
&&
\mbox{GL-type} \ : \ 
  g_{\gamma\pi\pi} \ \rightarrow\ 0 \,,
\nonumber\\
&&
\mbox{VM-type} \ : \ 
  g_{\gamma\pi\pi} \ \rightarrow\ \frac{1}{2} \,,
\nonumber\\
&&
\mbox{Hybrid-type} \ : \ 
  g_{\gamma\pi\pi} \ \rightarrow\ \frac{1}{3} \,.
\end{eqnarray}
In the GL-type $\zeta$ goes to 1, then the direct $\gamma\pi\pi$
goes to zero, which implies that the VD is sufficient.
On the other hand, in the VM-type the direct $\gamma\pi\pi$ 
approaches $1/2$, i.e., the VD is violated by $50\%$, similarly
to the case of the VM~\cite{HY:PRep,HS:VD}.
The hybrid type also gives about $33\%$ violation of the VD 
since the direct $\gamma\pi\pi$ comes to be $1/3$. 
This strongly affects to the understanding of the experimental data
on dilepton productions based on the dropping $\rho$
as recently pointed out in Ref.~\cite{Brown:2005ka-kb}.
An analysis of the spectral functions taking into
account of the large violation of the VD is much interesting.

Several comments are in order:

In this paper, we studied the chiral phase transition
with the first and second Weinberg sum rules kept satisfied.
Near the chiral restoration point
in QCD with a large number of massless flavors 
(the large $N_f$ QCD~\cite{largeNf,Appelquist:1996dq}),
the anomalous dimension $\gamma_m$ of 
$\langle \bar{q}q \rangle$ becomes close to $1$, 
$\gamma_m \sim 1$.
(See, e.g., Refs.~\cite{Appelquist:1996dq,HKY}.)
This 
results that
the term proportional to $\langle \bar{q}q \rangle^2$
in the current correlators behaves as 
$\sim 1/Q^4$~\cite{walking},
and hence the second Weinberg sum rule is not satisfied.
Thus, the condition in 
Eq.~(\ref{match rest}) is not appropriate for studying the
chiral symmetry restoration in the large $N_f$ QCD.

For studying the chiral phase transition at finite temperature
using the GHLS, which is relevant to the RHIC experiment,
we have to work explicitly under the existence of hot matter
as done for the HLS/VM in Refs.~\cite{HS:VM,HS:VD,PiV}.  
Since a typical energy scale characterizing the chiral symmetry
restoration, $T \sim 180$\,MeV, is much smaller than the momentum
scale $Q^2$ in the OPE,
we expect that the first and second Weinberg sum rules 
hold near the chiral phase transition point at finite temperature,
and that the matching condition in Eq.~(\ref{match rest}) is applicable.
Then, when the analysis in this paper is applicable to the
chiral phase transition at finite temperature, we expect that 
the matching will produce the temperature dependence of 
the GHLS gauge coupling similar to the one of the HLS gauge coupling
(intrinsic temperature dependence),
which was essential to describe the dropping $\rho$ occurring 
in the VM, and that the dropping $\rho$ and $A_1$ masses are
suitably described within the GHLS.

In the GHLS sector,
the theory space locality is broken due to ${\cal O}(p^4)$ terms 
and we do not have the non-renormalization of the second Weinberg 
sum rule at one-loop.
The relation between the higher order terms which generate a violation 
of the sum rule and the chiral symmetry restoration will be
elucidated by forthcoming analysis.

In the case of the HLS, the extrapolation of the ChPT
from large $N_c$ QCD to the real-life QCD together with the
Wilsonian matching reproduces several physical quantities
in remarkable agreement with experiments~\cite{HY:PRep,HY:WM}.
The GHLS at the leading order with a suitable parameter 
choice was shown~\cite{BKY:NPB,BKY:PRep} to explain several 
phenomenological facts such as the successful current algebra 
relations $m_{A_1}^2 = 2 m_\rho^2$ and $g_{A_1}=g_\rho$
~\cite{SVZ,yamawaki}.
Furthermore, inclusion of appropriate higher order terms 
gives predictions on the widths of $A_1 \rightarrow \rho \pi$
and $A_1 \rightarrow \gamma \pi$ in good agreement with
experiments~\cite{BFY:GHLS,BKY:PRep}.
Then, we believe that the ChPT with GHLS incorporated by
a suitable matching procedure will describe the low-energy
phenomenology of real-life QCD.
In the present analysis, 
we focused on the phase structure of the GHLS theory.
It is interesting to study the physical quantities
such as the $\rho$-$\pi$-$\pi$ coupling and $\rho$-$\gamma$
mixing through the matching procedure based on the
ChPT with GHLS developed in this paper.
We leave the analysis in the future publications.

One might think that 
the scalar mesons should be included,
since several analysis~\cite{scalars}
shows that they are
lighter than the vector mesons in real-life QCD.
For example, the analysis in Ref.~\cite{HSS} shows that
the mass of sigma meson is about 560\,MeV, 
which is definitely lighter than
the $\rho$ meson, $m_\rho = 770$\,MeV.
In this paper,
we did not include scalar meson as a dynamical degree of freedom,
assuming that the light $\sigma$ meson is made of two quarks and two
anti-quarks~\cite{Jaffe,BFSS} and irrelevant to the present analysis.
It is possible that
another scalar meson made of $q \bar{q}$ will appear
quite near the chiral restoration point and we may have to
take the effects into account.
Inclusion of the light scalar meson quite
near the critical point
may generate the quadratic divergence
and change the present RGEs.


\section*{Acknowledgment}

We are grateful to Yoshimasa Hidaka, Mannque Rho
Masaharu Tanabashi and Koichi Yamawaki 
for useful discussions and comments.
The work of M.H. 
is supported in part by the Daiko Foundation \#9099, 
the 21st Century
COE Program of Nagoya University provided by Japan Society for the
Promotion of Science (15COEG01), and the JSPS Grant-in-Aid for
Scientific Research (c) (2) 16540241.
The work of C.S. is supported in part by the Virtual Institute
of the Helmholtz Association under the grant No. VH-VI-041.


\section*{Note added}

After the completion of this work,
we were aware of a similar analysis by Hidaka, Morimatsu and 
Ohtani~\cite{HMO}.


\appendix

\setcounter{section}{0}
\renewcommand{\thesection}{\Alph{section}}
\setcounter{equation}{0}
\renewcommand{\theequation}{\Alph{section}.\arabic{equation}}


\section{Background Field Gauge}
\label{app:BFG}

In this appendix, we perform the quantization of the GHLS theory
in the background field gauge.

In order that the combinations $(\phi_R \pm \phi_L)/2$
belong to the parity-eigenstates,
we insert a local symmetry $H^\prime = [SU(N_f)_V]_{\rm local}$
by dividing $\xi_M$ into two parts as
\begin{equation}
\xi_M = \xi_{ML}^\dagger\cdot\xi_{MR}\,.
\end{equation}
Two variables $\xi_{ML}$ and $\xi_{MR}$ transform as
\begin{eqnarray}
&&
\xi_{ML} \to h^\prime\xi_{ML}h_L^\dagger\,,
\nonumber\\
&&
\xi_{MR} \to h^\prime\xi_{MR}h_R^\dagger\,,
\end{eqnarray}
where $h^\prime \in [SU(N_f)_V]_{\rm local}$.
Accordingly we introduce the background fields $\bar{\xi}$
and the quantum fields $\check{\xi}$ as
  \begin{eqnarray}
  &&
  \xi_L 
  = \bar{\xi}_{pL}^\dag\check{\xi}_L\bar{\xi}_{pL}\bar{\xi}_L\,,
  \nonumber\\
  &&
  \xi_R 
  = \bar{\xi}_{pR}^\dagger\check{\xi}_R\bar{\xi}_{pR}\bar{\xi}_R\,,
  \nonumber\\
  &&
  \xi_M
  = \bar{\xi}_{pL}^\dag \check{\xi}_M\bar{\xi}_{pR}\,.
  \end{eqnarray}
The transformation properties of $\bar{\xi}$ and $\check{\xi}$
are given by
\begin{eqnarray}
&&
\bar{\xi}_L \to h_L\bar{\xi}_L g_L^\dagger\,,
\qquad
\bar{\xi}_R \to h_R\bar{\xi}_R g_R^\dagger\,,
\nonumber\\
&&
\bar{\xi}_{pL} \to h^\prime \bar{\xi}_{pL} h_L^{\dagger}\,,
\quad
\bar{\xi}_{pR} \to h^\prime\bar{\xi}_{pR}h_R^\dagger\,,
\nonumber\\
&&
\check{\xi}_L \to h^\prime\check{\xi}_L h^{\prime\dagger}\,,
\qquad
\check{\xi}_R \to h^\prime\check{\xi}_R h^{\prime\dagger}\,,
\nonumber\\
&&
\check{\xi}_M \to h^\prime\check{\xi}_M h^{\prime\dagger}\,,
\end{eqnarray}
Then all the quantum fields transform homogeneously
under the background gauge transformation:
\begin{equation}
(\check{\pi}, \check{\sigma}, \check{q}) \to 
h^\prime(\check{\pi}, \check{\sigma}, \check{q})h^{\prime\dagger}\,.
\end{equation}
The background and quantum fields of the GHLS gauge bosons are
introduced as follows:
\begin{eqnarray}
&&
L_\mu = \bar{L}_\mu 
{}+ g\bar{\xi}_{pL}^\dagger\check{L}_\mu\bar{\xi}_{pL}\,,
\nonumber\\
&&
R_\mu = \bar{R}_\mu 
{}+ g\bar{\xi}_{pR}^\dagger\check{R}_\mu\bar{\xi}_{pR}\,,
\end{eqnarray}
which transform as
\begin{eqnarray}
&&
\bar{L}_\mu \to h_L\bar{L}_\mu h_L^\dagger 
{}+ ih_L\partial_\mu h_L^\dagger\,,
\nonumber\\
&&
\bar{R}_\mu \to h_R\bar{R}_\mu h_R^\dagger 
{}+ ih_R\partial_\mu h_R^\dagger\,,
\nonumber\\
&&
\check{L}_\mu \to h^\prime\check{L}_\mu h^{\prime\dagger}\,,
\quad
\check{R}_\mu \to h^\prime\check{R}_\mu h^{\prime\dagger}\,.
\end{eqnarray}
The covariant derivatives acting on $\bar{\xi}$ and $\check{\xi}$
are expressed as
\begin{eqnarray}
&&
\bar{D}_\mu\bar{\xi}_{L} = \partial_\mu\bar{\xi}_{L}
{}- i\bar{L}_\mu\bar{\xi}_{L}
{}+ i\bar{\xi}_{L} {\mathcal L}_\mu\,
\nonumber\\
&&
\bar{D}_\mu\bar{\xi}_{R} = \partial_\mu\bar{\xi}_{R}
{}- i\bar{R}_\mu\bar{\xi}_{R}
{}+ i\bar{\xi}_{R} {\mathcal R}_\mu\,
\nonumber\\
&&
\bar{D}_\mu\bar{\xi}_{pL} = \partial_\mu\bar{\xi}_{pL}
{}- i\bar{V}_\mu^\prime\bar{\xi}_{pL}
{}+ i\bar{\xi}_{pL} \bar{L}_\mu\,
\nonumber\\
&&
\bar{D}_\mu\bar{\xi}_{pR} = \partial_\mu\bar{\xi}_{pR}
{}- i\bar{V}_\mu^\prime\bar{\xi}_{pR}
{}+ i\bar{\xi}_{pR}\bar{R}_\mu \,,
\nonumber\\
&&
\bar{D}_\mu\check{\xi}_{L,R,M} = \partial_\mu\check{\xi}_{L,R,M} 
{}- i[\bar{V}_\mu^\prime, \check{\xi}_{L,R,M}]\,,
\end{eqnarray}
where $\bar{V}_\mu^\prime$ is the background gauge field
corresponding to the $H^\prime$ symmetry.

It is convenient to define the back ground fields
associated with the external gauge fields as follows:
\begin{eqnarray}
\bar{\mathcal V}_\parallel^\mu
&=&
\frac{1}{2i}
\left[
  \bar{\xi}_{pR} 
    \partial^\mu \bar{\xi}_R \cdot \bar{\xi}_R^\dag
  \bar{\xi}_{pR}^\dag
  +
  \bar{\xi}_{pL} 
    \partial^\mu \bar{\xi}_L \cdot \bar{\xi}_L^\dag
  \bar{\xi}_{pL}^\dag
\right]
\nonumber\\
&& {}
+
\frac{1}{2}
\left[
  \bar{\xi}_{pR} 
    \bar{\xi}_{R} {\mathcal R}^\mu \bar{\xi}_R^\dag
  \bar{\xi}_{pR}^\dag
  +
  \bar{\xi}_{pL} 
    \bar{\xi}_{L} {\mathcal L}^\mu \bar{\xi}_L^\dag
  \bar{\xi}_{pL}^\dag
\right]
\ ,
\nonumber\\
\bar{\mathcal A}_\perp^\mu
&=&
\frac{1}{2i}
\left[
  \bar{\xi}_{pR} 
    \partial^\mu \bar{\xi}_R \cdot \bar{\xi}_R^\dag
  \bar{\xi}_{pR}^\dag
  -
  \bar{\xi}_{pL} 
    \partial^\mu \bar{\xi}_L \cdot \bar{\xi}_L^\dag
  \bar{\xi}_{pL}^\dag
\right]
\nonumber\\
&& {}
+
\frac{1}{2}
\left[
  \bar{\xi}_{pR} 
    \bar{\xi}_{R} {\mathcal R}^\mu \bar{\xi}_R^\dag
  \bar{\xi}_{pR}^\dag
  -
  \bar{\xi}_{pL} 
    \bar{\xi}_{L} {\mathcal L}^\mu \bar{\xi}_L^\dag
  \bar{\xi}_{pL}^\dag
\right]
\ .
\nonumber\\
\end{eqnarray}
Furthermore, we use
\begin{eqnarray}
&&
\bar{V}_\mu
= \frac{1}{2} \left[
  \bar{\xi}_{pR}\bar{R}_\mu\bar{\xi}_{pR}^\dagger
  {} + \bar{\xi}_{pL}\bar{L}_\mu\bar{\xi}_{pL}^\dag
\right]\,,
\nonumber\\
&&
\bar{A}_\mu
= \frac{1}{2} \left[
  \bar{\xi}_{pR}\bar{R}_\mu\bar{\xi}_{pR}^\dagger
  {} - \bar{\xi}_{pL}\bar{L}_\mu\bar{\xi}_{pL}^\dag
\right]\,,
\label{sub:2}
\end{eqnarray}
and
\begin{eqnarray}
\bar{\mathcal V}_M^\mu
&=&
\frac{1}{2i}
\left[
  \partial^\mu \bar{\xi}_{pR} \cdot \bar{\xi}_{pR}^\dag
  +
  \partial^\mu \bar{\xi}_{pL} \cdot \bar{\xi}_{pL}^\dag
\right]
\,,
\nonumber\\
\bar{\mathcal A}_M^\mu
&=&
\frac{1}{2i}
\left[
  \partial^\mu \bar{\xi}_{pR} \cdot \bar{\xi}_{pR}^\dag
  -
  \partial^\mu \bar{\xi}_{pL} \cdot \bar{\xi}_{pL}^\dag
\right]
\,.
\end{eqnarray}

We fix the background field gauge as
\begin{eqnarray}
{\cal L}_{\rm GF}^{(V + A)}
=
- \frac{1}{\alpha} \, \mbox{tr} \left[ F_V^2 + F_A^2 \right]
\ ,
\label{GF}
\end{eqnarray}
where $\alpha$ is the gauge fixing parameter,
and $F_V$ and $F_A$ are defined as
\begin{eqnarray}
F_V &=&
 \bar{D}^\mu \check{V}_\mu + \alpha g F_\sigma^2 \check{\phi}_\sigma
 + i \left[ \bar{V}^{\prime\mu} - \bar{V}^\mu 
          - \bar{\mathcal V}_M^\mu \,,\, \check{V}_\mu \right]
\nonumber\\
&& {} 
 - i \left[ \bar{A}^\mu + \bar{\mathcal A}_M^\mu \,,\, 
            \check{A}_\mu \right]
\,, 
\nonumber\\
F_A &=&
 \bar{D}^\mu \check{A}_\mu - \alpha g F_q^2 \check{\phi}_q
 + i \left[ \bar{V}^{\prime\mu} - \bar{V}^\mu 
          - \bar{\mathcal V}_M^\mu \,,\, \check{A}_\mu \right]
\nonumber\\
&& {} 
 - i \left[ \bar{A}^\mu + \bar{\mathcal A}_M^\mu \,,\, 
            \check{V}_\mu \right]
\,, 
\end{eqnarray}
with 
$\bar{D}^\mu$ denoting the covariant derivative
in terms of the background fields:
\begin{eqnarray}
&&
\bar{D}_\mu\check{V}^\nu = 
\partial_\mu\check{V}^\nu - i[\bar{V}_\mu^\prime, \check{V}^\nu ]\,,
\nonumber\\
&&
\bar{D}_\mu\check{A}^\nu = 
\partial_\mu\check{A}^\nu - i[\bar{V}_\mu^\prime, \check{A}^\nu ]\,.
\end{eqnarray}

The FP ghost Lagrangian 
associated with ${\cal L}_{\rm GF}$ is given by
\begin{eqnarray}
&&
{\cal L}_{\rm FP}^{(V + A)}
= 2i\mbox{tr}\Bigl[ \bar{C}_V \Bigl\{ \bar{D}_\mu
\Bigl( \bar{D}^\mu C_V + \alpha g^2 F_\sigma^2 C_V  
\nonumber\\
&&
{}- i[\bar{V}^{\prime\mu} - \bar{V}^\mu
{}- \bar{\cal V}_M^\mu, C_V] - i[\bar{A}^\mu 
{}+ \bar{\cal A}_M^\mu, C_A]
\Bigr)
\Bigr\}
\Bigr]
\nonumber\\
&&
{}+ 2i\mbox{tr}\Bigl[ \bar{C}_A \Bigl\{ \bar{D}_\mu
\Bigl( \bar{D}^\mu C_A + \alpha g^2 F_q^2 C_A  
\nonumber\\
&&\qquad
{}- i[\bar{V}^{\prime\mu} - \bar{V}^\mu
{}- \bar{\cal V}_M^\mu, C_A] - i[\bar{A}^\mu 
{}+ \bar{\cal A}_M^\mu, C_V]
\Bigr)
\Bigr\}
\Bigr]
\nonumber\\
&&
{}- 2\mbox{tr}\Bigl[\Bigl\{ [\bar{V}^{\prime\mu} - \bar{V}^\mu
{}- \bar{\cal V}_M^\mu, \bar{C}_V] + [\bar{A}^\mu 
{}+ \bar{\cal A}_M^\mu, \bar{C}_A]
\Bigr\}
\nonumber\\
&&\quad\times
\Bigl\{\bar{D}_\mu C_V - i[\bar{V}^{\prime\mu} - \bar{V}^\mu
{}- \bar{\cal V}_M^\mu, C_V] 
\nonumber\\
&&\qquad
{}- i[\bar{A}^\mu + \bar{\cal A}_M^\mu, C_A]
\Bigr\}
\Bigr]
\nonumber\\
&&
{}- 2\mbox{tr}\Bigl[\Bigl\{ [\bar{V}^{\prime\mu} - \bar{V}^\mu
{}- \bar{\cal V}_M^\mu, \bar{C}_A] + [\bar{A}^\mu 
{}+ \bar{\cal A}_M^\mu, \bar{C}_V]
\Bigr\}
\nonumber\\
&&\quad\times
\Bigl\{\bar{D}_\mu C_A - i[\bar{V}^{\prime\mu} - \bar{V}^\mu
{}- \bar{\cal V}_M^\mu, C_A] 
\nonumber\\
&&\qquad
{}- i[\bar{A}^\mu + \bar{\cal A}_M^\mu, C_V]
\Bigr\}
\Bigr]
{}+ \cdots\,,
\label{FP}
\end{eqnarray}
where ellipses stand for the terms including at least three
quantum fields.

Finally, we should eliminate 
the redundant $H^\prime$ symmetry.
This can be done by relating 
$\bar{V}_\mu^\prime$ to other background fields.
In this paper we take
\begin{equation}
\bar{V}^{^\prime\mu}
= \bar{\cal V}_{M}^\mu + \bar{V}^\mu\,.
\label{H'-GF}
\end{equation}
and set 
$\bar{\xi}_{pL}=\bar{\xi}_{pR}\equiv \bar{\xi}_p$.
We should note that
the background field $\bar{V}_\mu^\prime$ appears
only in ${\cal L}_{\rm GF}+{\cal L}_{\rm FP}$ given
in Eqs.~(\ref{GF}) and (\ref{FP}),
and is not included in the Lagrangian in 
Eq.~(\ref{lag p^2}).


\section{Quantum Corrections}
\label{app:QC}

In this appendix, we list the quantum corrections to
the two-point functions and the RGEs at one loop.
In the present analysis, we adopt the 't~Hooft-Feynman gauge 
by taking $\alpha=1$ in Eqs.~(\ref{GF}) and (\ref{FP}).

For expressing the quantum corrections
in simple forms it is convenient to define
the following Feynman integrals:
\begin{eqnarray}
&&
A_0(M)
= \int \frac{d^n k}{i(2\pi)^4}\frac{1}{M^2 - k^2}\,,
\nonumber\\
&&
\nonumber\\
&&
B_0(p;M_1,M_2)
\nonumber\\
&&\quad
= \int \frac{d^n k}{i(2\pi)^4}
  \frac{1}{[M_1^2-k^2][M_2^2-(k-p)^2]}\,, 
\nonumber\\
&&
B^{\mu\nu}(p;M_1,M_2)
\nonumber\\
&&\quad
= \int \frac{d^n k}{i(2\pi)^4}
  \frac{(2k-p)^\mu (2k-p)^\nu }{[M_1^2-k^2][M_2^2-(k-p)^2]}\,.
\end{eqnarray}
We take into account
the quadratic as well as the logarithmic divergences
following Eqs.~(\ref{quad}) and (\ref{log}),
which preserve the chiral symmetry.
Here we summarize the divergent parts of the above
Feynman integrals:
\begin{eqnarray}
&&
\left. A_0(M) \right\vert_{\rm div}
=
\frac{\Lambda^2}{(4\pi)^2} - \frac{M^2}{(4\pi)^2} \, \ln \Lambda^2
\ ,
\nonumber\\
&&
\left. B_0(p;M_1,M_2) \right\vert_{\rm div}
=
\frac{1}{(4\pi)^2} \, \ln \Lambda^2 \ ,
\nonumber\\
&&
\left. B^{\mu\nu}(p;M_1,M_2) \right\vert_{\rm div}
\nonumber\\
&& \quad
=
- g^{\mu\nu} \frac{1}{(4\pi)^2}
\left[ 
  2 \Lambda^2 - \left( M_1^2 + M_2^2 \right) \ln \Lambda^2
\right]
\nonumber\\
&& \qquad
 {}- \left( p^2 g^{\mu\nu} - p^\mu p^\nu \right)
   \frac{1}{3(4\pi)^2} \, \ln \Lambda^2
\ .
\end{eqnarray}

\begin{figure}
 \begin{center}
  \includegraphics[width = 8cm]{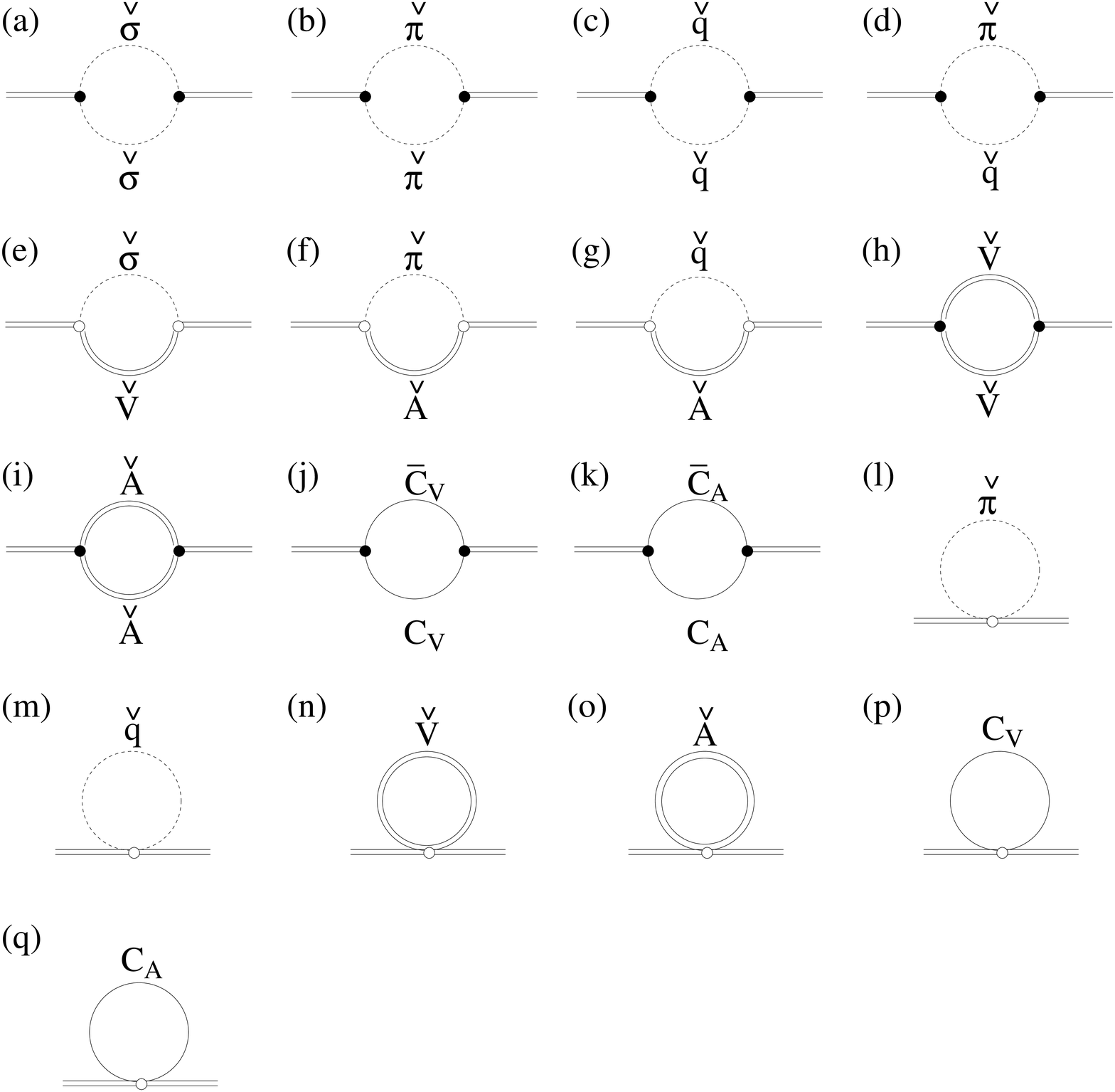}
 \end{center}
 \caption{
 Diagrams for contributions to $\Pi_{\bar{V}\bar{V}}^{\mu\nu}$
 at one loop. The circle $(\circ)$ denotes the momentum-independent 
 vertex and the dot $(\bullet)$ denotes the momentum-dependent vertex.
 }
 \label{fig:VV}
\end{figure}
The quantum corrections to $\Pi_{\bar{V}\bar{V}}^{\mu\nu}$
generated by the loop diagrams shown in Fig.~\ref{fig:VV}
are given by
\begin{eqnarray}
&&
\Pi_{\bar{V}\bar{V}}^{{\rm (a)}\mu\nu}
= \frac{N_f}{8}B^{\mu\nu}(p;M_\rho,M_\rho)\,,
\nonumber\\
&&
\Pi_{\bar{V}\bar{V}}^{{\rm (b)}\mu\nu}
= \frac{N_f}{8}\Bigl( \frac{F}{F_\pi} \Bigr)^4
  a^2 (1-\zeta)^2 (1+\zeta)^2 B^{\mu\nu}(p;0,0)\,,
\nonumber\\
&&
\Pi_{\bar{V}\bar{V}}^{{\rm (c)}\mu\nu}
= \frac{N_f}{8}\Bigl(\frac{F}{F_q}\Bigr)^4
  [a - 2(b+c)]^2 B^{\mu\nu}(p;M_{A_1},M_{A_1})\,,
\nonumber\\
&&
\Pi_{\bar{V}\bar{V}}^{{\rm (d)}\mu\nu}
= \frac{N_f}{4}\Bigl( \frac{F^2}{F_\pi F_q} \Bigr)^2
  (a\zeta - b)^2  
B^{\mu\nu}(p;M_{A_1},0)\,,
\nonumber\\
&&
\Pi_{\bar{V}\bar{V}}^{{\rm (e)}\mu\nu}
= - N_f M_\rho^2 g^{\mu\nu} B_0(p;M_\rho,M_\rho)\,,
\nonumber\\
&&
\Pi_{\bar{V}\bar{V}}^{{\rm (f)}\mu\nu}
= - N_f \Bigl( \frac{F}{F_\pi} \Bigr)^2 g^2 F^2
 (a\zeta - b)^2 
\nonumber\\
&&\qquad\qquad
\times g^{\mu\nu}B_0(p;M_{A_1},0)\,,
\nonumber\\
&&
\Pi_{\bar{V}\bar{V}}^{{\rm (g)}\mu\nu}
= - N_f \Bigl( \frac{F}{F_q}\Bigr)^2 g^2 a^2 F^2 g^{\mu\nu}
  B_0(p;M_{A_1},M_{A_1})\,,
\nonumber\\
&&
\Pi_{\bar{V}\bar{V}}^{{\rm (h)}\mu\nu}
= \frac{N_f}{2}\Bigl[ n B^{\mu\nu}(p;M_\rho,M_\rho)
\nonumber\\
&&\qquad\qquad
{}+ 8(p^2 g^{\mu\nu} - p^\mu p^\nu )B_0(p;M_\rho,M_\rho)\Bigr]\,,
\nonumber\\
&&
\Pi_{\bar{V}\bar{V}}^{{\rm (i)}\mu\nu}
= \frac{N_f}{2}\Bigl[ n B^{\mu\nu}(p;M_{A_1},M_{A_1})
\nonumber\\
&&\qquad\qquad
{}+ 8(p^2 g^{\mu\nu} - p^\mu p^\nu )B_0(p;M_{A_1},M_{A_1})\Bigr]\,,
\nonumber\\
&&
\Pi_{\bar{V}\bar{V}}^{{\rm (j)}\mu\nu}
= - N_f B^{\mu\nu}(p;M_\rho,M_\rho)\,,
\nonumber\\
&&
\Pi_{\bar{V}\bar{V}}^{{\rm (k)}\mu\nu}
= - N_f B^{\mu\nu}(p;M_{A_1},M_{A_1})\,,
\nonumber\\
&&
\Pi_{\bar{V}\bar{V}}^{{\rm (l)}\mu\nu}
= - N_f \Bigl( \frac{F}{F_\pi}\Bigr)^2 \zeta(a\zeta - b)
g^{\mu\nu} A_0(0)\,,
\nonumber\\
&&
\Pi_{\bar{V}\bar{V}}^{{\rm (m)}\mu\nu}
= N_f \Bigl(\frac{F}{F_q}\Bigr)^4 (a-b-c)^2
 g^{\mu\nu} A_0(M_{A_1})\,,
\nonumber\\
&&
\Pi_{\bar{V}\bar{V}}^{{\rm (n)}\mu\nu}
= N_f n\, g^{\mu\nu} A_0(M_\rho)\,,
\nonumber\\
&&
\Pi_{\bar{V}\bar{V}}^{{\rm (o)}\mu\nu}
=  N_f n\, g^{\mu\nu} A_0(M_{A_1})\,,
\nonumber\\
&&
\Pi_{\bar{V}\bar{V}}^{{\rm (p)}\mu\nu}
= - 2N_f g^{\mu\nu} A_0(M_\rho)\,,
\nonumber\\
&&
\Pi_{\bar{V}\bar{V}}^{{\rm (q)}\mu\nu}
= - 2N_f g^{\mu\nu} A_0(M_{A_1})\,,
\label{eq:appB:1}
\end{eqnarray}
where $n$ denotes the dimension of the spacetime.
\begin{figure}
 \begin{center}
  \includegraphics[width = 8cm]{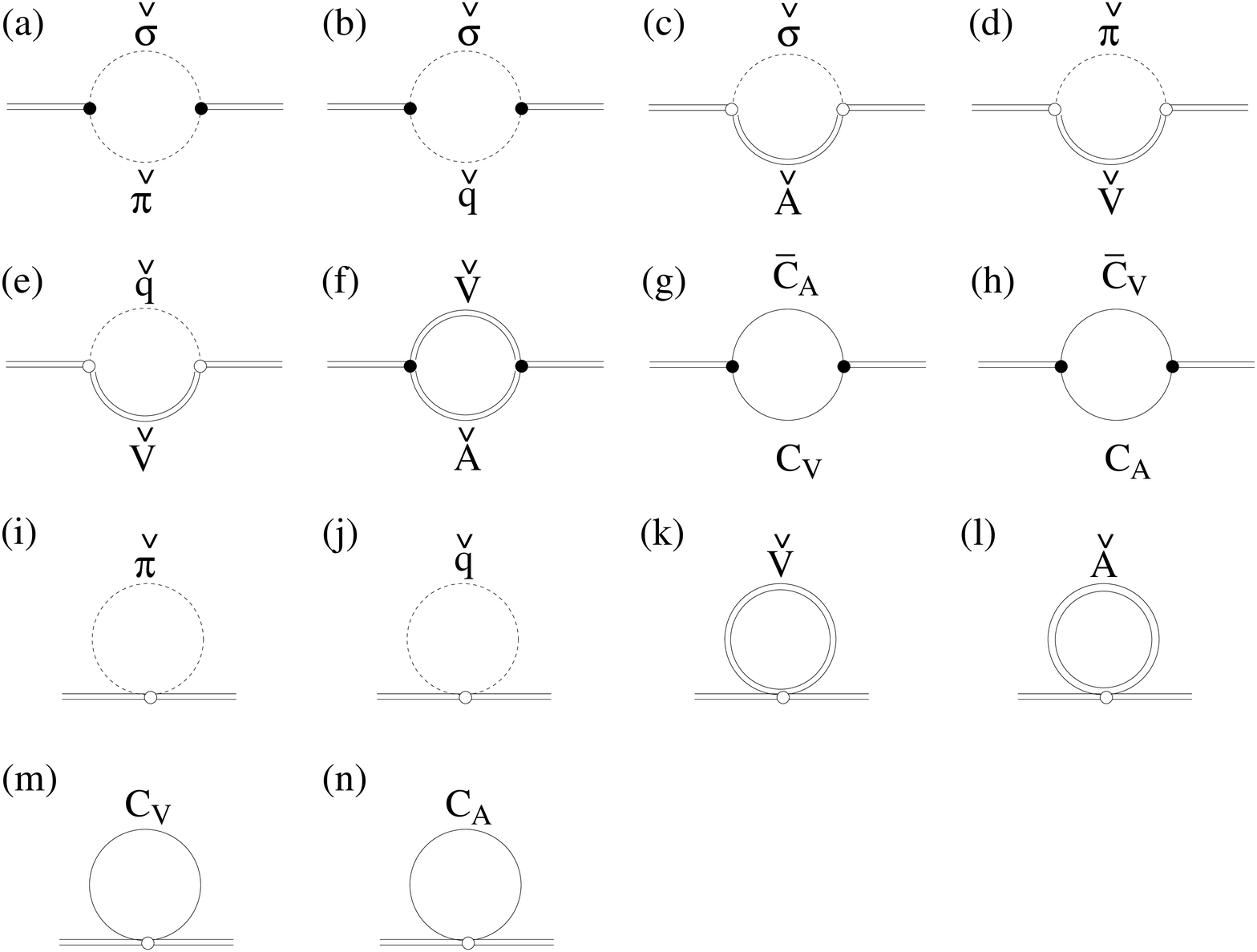}
 \end{center}
 \caption{
 Diagrams for contributions to $\Pi_{\bar{A}\bar{A}}^{\mu\nu}$
 at one loop.
 }
 \label{fig:AA}
\end{figure}
We show the one-loop diagrams to contribute to 
$\Pi_{\bar{A}\bar{A}}^{\mu\nu}$ in Fig.~\ref{fig:AA}.
We obtain the quantum corrections as
\begin{eqnarray}
&&
\Pi_{\bar{A}\bar{A}}^{{\rm (a)}\mu\nu}
= \frac{N_f}{4}\Bigl(\frac{F^2}{F_\sigma F_\pi} \Bigr)^2
 (a\zeta - b)^2 
B^{\mu\nu}(p;M_\rho,0)\,,
\nonumber\\
&&
\Pi_{\bar{A}\bar{A}}^{{\rm (b)}\mu\nu}
= \frac{N_f}{4} a^2 \Bigl(\frac{F^2}{F_\sigma F_q} \Bigr)^2
  B^{\mu\nu}(p;M_\rho,M_{A_1})\,,
\nonumber\\
&&
\Pi_{\bar{A}\bar{A}}^{{\rm (c)}\mu\nu}
= - N_f M_\rho^2 g^{\mu\nu} B_0(p;M_\rho,M_{A_1})\,,
\nonumber\\
&&
\Pi_{\bar{A}\bar{A}}^{{\rm (d)}\mu\nu}
= - N_f \Bigl( \frac{F}{F_\pi}\Bigr)^2 g^2 F^2 \zeta^2
  (a-b-c)^2 
\nonumber\\
&&\qquad\qquad
\times g^{\mu\nu} B_0(p;M_\rho,0)\,,
\nonumber\\
&&
\Pi_{\bar{A}\bar{A}}^{{\rm (e)}\mu\nu}
= - N_f \Bigl( \frac{F}{F_q}\Bigr)^2 g^2 F^2
  [a - 2(b+c)]^2 
\nonumber\\
&&\qquad\qquad
\times g^{\mu\nu} B_0(p;M_\rho,M_{A_1})\,,
\nonumber\\
&&
\Pi_{\bar{A}\bar{A}}^{{\rm (f)}\mu\nu}
= N_f \Bigl[ n B^{\mu\nu}(p;M_\rho,M_{A_1})
\nonumber\\
&&\qquad\qquad
{}+ 8(p^2 g^{\mu\nu} - p^\mu p^\nu )B_0(p;M_\rho,M_{A_1})\Bigr]\,,
\nonumber\\
&&
\Pi_{\bar{A}\bar{A}}^{{\rm (g)}\mu\nu}
= - N_f B^{\mu\nu}(p;M_\rho,M_{A_1})\,,
\nonumber\\
&&
\Pi_{\bar{A}\bar{A}}^{{\rm (h)}\mu\nu}
= - N_f B^{\mu\nu}(p;M_\rho,M_{A_1})\,,
\nonumber\\
&&
\Pi_{\bar{A}\bar{A}}^{{\rm (i)}\mu\nu}
= N_f \Bigl(\frac{F}{F_\pi}\Bigr)^2 (a-b-c)\zeta^2
  g^{\mu\nu} A_0(0)\,,
\nonumber\\
&&
\Pi_{\bar{A}\bar{A}}^{{\rm (j)}\mu\nu}
= N_f \Bigl(\frac{F}{F_q}\Bigr)^2 (a-b-c)
  g^{\mu\nu} A_0(M_{A_1})\,,
\nonumber\\
&&
\Pi_{\bar{A}\bar{A}}^{{\rm (k)}\mu\nu}
= N_f n\, g^{\mu\nu} A_0(M_\rho)\,,
\nonumber\\
&&
\Pi_{\bar{A}\bar{A}}^{{\rm (l)}\mu\nu}
= N_f n\, g^{\mu\nu} A_0(M_{A_1})\,,
\nonumber\\
&&
\Pi_{\bar{A}\bar{A}}^{{\rm (m)}\mu\nu}
= - 2N_f g^{\mu\nu} A_0(M_\rho)\,,
\nonumber\\
&&
\Pi_{\bar{A}\bar{A}}^{{\rm (n)}\mu\nu}
= - 2N_f g^{\mu\nu} A_0(M_{A_1})\,.
\end{eqnarray}
\begin{figure}
 \begin{center}
  \includegraphics[width = 8cm]{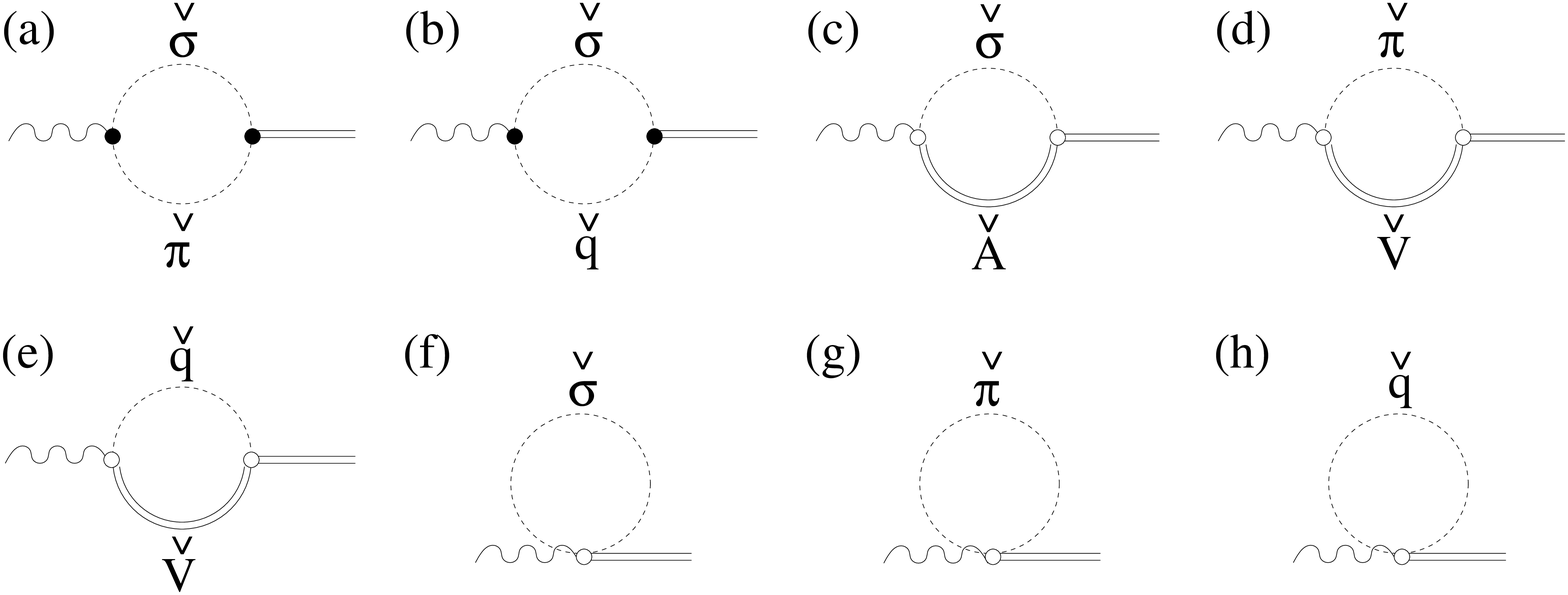}
 \end{center}
 \caption{
 Diagrams for contributions to 
 $\Pi_{\bar{\cal A}_\perp\bar{A}}^{\mu\nu}$ at one loop.
 }
 \label{fig:perpA}
\end{figure}
{}From the diagrams shown in Fig.~\ref{fig:perpA},
we obtain
\begin{eqnarray}
&&
\Pi_{\bar{\cal A}_\perp\bar{A}}^{{\rm (a)}\mu\nu}
= - \frac{N_f}{4}\Bigl(\frac{F^2}{F_\sigma F_\pi}\Bigr)^2
  (a - b\zeta) 
\nonumber\\
&&\qquad\qquad
\times (a\zeta - b) B^{\mu\nu}(p;M_\rho,0)\,,
\nonumber\\
&&
\Pi_{\bar{\cal A}_\perp\bar{A}}^{{\rm (b)}\mu\nu}
= \frac{N_f}{4}\Bigl(\frac{F^2}{F_\sigma F_q}\Bigr)^2
  a b B^{\mu\nu}(p;M_\rho,M_{A_1})\,,
\nonumber\\
&&
\Pi_{\bar{\cal A}_\perp\bar{A}}^{{\rm (c)}\mu\nu}
= N_f \Bigl(\frac{F}{F_\sigma}\Bigr)^2 g^2 F^2 a b
  g^{\mu\nu} B_0(p;M_\rho,M_{A_1})\,,
\nonumber\\
&&
\Pi_{\bar{\cal A}_\perp\bar{A}}^{{\rm (d)}\mu\nu}
= N_f \Bigl(\frac{F}{F_\pi}\Bigr)^2 g^2 F^2
  \zeta(a - b\zeta)
\nonumber\\
&&\qquad\qquad
\times (a-b-c) g^{\mu\nu} B_0(p;M_\rho,0)\,,
\nonumber\\
&&
\Pi_{\bar{\cal A}_\perp\bar{A}}^{{\rm (e)}\mu\nu}
= - N_f \Bigl(\frac{F}{F_q}\Bigr)^2 g^2 F^2
  b [a - 2(b+c)] 
\nonumber\\
&&\qquad\qquad
\times g^{\mu\nu} B_0(p;M_\rho,M_{A_1})\,,
\nonumber\\
&&
\Pi_{\bar{\cal A}_\perp\bar{A}}^{{\rm (f)}\mu\nu}
= \frac{N_f}{2}\Bigl(\frac{F}{F_\sigma}\Bigr)^2 b
  g^{\mu\nu} A_0(M_\rho)\,,
\nonumber\\
&&
\Pi_{\bar{\cal A}_\perp\bar{A}}^{{\rm (g)}\mu\nu}
= N_f \Bigl(\frac{F}{F_\pi}\Bigr)^2
  \Bigl[ - a\zeta + b\Bigl( 1 - \frac{1}{2}(1-\zeta^2)\Bigr) \Bigr]
\nonumber\\
&&\qquad\qquad
\times  g^{\mu\nu}A_0(0)\,,
\nonumber\\
&&
\Pi_{\bar{\cal A}_\perp\bar{A}}^{{\rm (h)}\mu\nu}
= \frac{N_f}{2}\Bigl(\frac{F}{F_q}\Bigr)^2 b
  g^{\mu\nu} A_0(M_{A_1})\,.
\end{eqnarray}
\begin{figure}
 \begin{center}
  \includegraphics[width = 8cm]{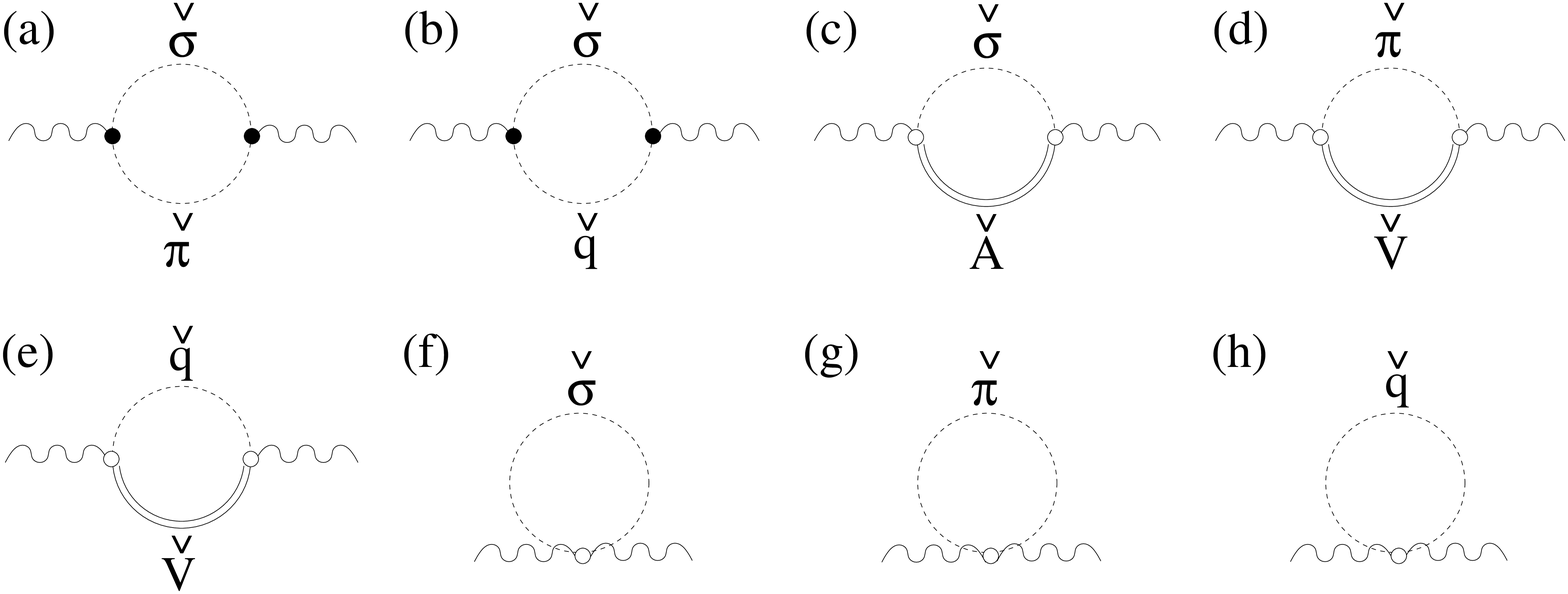}
 \end{center}
 \caption{
 Diagrams for contributions to 
 $\Pi_{\bar{\cal A}_{M}\bar{\cal A}_\perp}^{\mu\nu}$
 at one loop.
 }
 \label{fig:Mperpperp}
\end{figure}
Finally we list the quantum corrections contributing to
$\Pi_{\bar{\cal A}_{M}\bar{\cal A}_\perp}^{\mu\nu}$ 
shown in Fig.~\ref{fig:Mperpperp}:
\begin{eqnarray}
&&
\Pi_{\bar{\cal A}_{M}\bar{\cal A}_\perp}^{{\rm (a)}\mu\nu}
= \frac{N_f}{4}\Bigl(\frac{F^2}{F_\sigma F_\pi}\Bigr)^2
  (1-\zeta) (a+b) (a - b\zeta) 
\nonumber\\
&&\qquad\qquad
\times B^{\mu\nu}(p;M_\rho,0)\,,
\nonumber\\
&&
\Pi_{\bar{\cal A}_{M}\bar{\cal A}_\perp}^{{\rm (b)}\mu\nu}
= \frac{N_f}{4}\Bigl(\frac{F^2}{F_\sigma F_q}\Bigr)^2
  b (a+b) B^{\mu\nu}(p;M_\rho,M_{A_1})\,,
\nonumber\\
&&
\Pi_{\bar{\cal A}_{M}\bar{\cal A}_\perp}^{{\rm (c)}\mu\nu}
= N_f \Bigl(\frac{F}{F_\sigma}\Bigr)^2 g^2 F^2
  b (a-b) 
\nonumber\\
&&\qquad\qquad
\times g^{\mu\nu} B_0(p;M_\rho,M_{A_1})\,,
\nonumber\\
&&
\Pi_{\bar{\cal A}_{M}\bar{\cal A}_\perp}^{{\rm (d)}\mu\nu}
= - N_f \Bigl(\frac{F}{F_\pi}\Bigr)^2 g^2 F^2 (a - b\zeta)
\nonumber\\
&&\qquad\qquad
\times  [a(1-\zeta) + c\zeta] g^{\mu\nu} B_0(p;M_\rho,0)\,,
\nonumber\\
&&
\Pi_{\bar{\cal A}_{M}\bar{\cal A}_\perp}^{{\rm (e)}\mu\nu}
= - N_f \Bigl(\frac{F}{F_q}\Bigr)^2 g^2 F^2
  b (a-b-2c) 
\nonumber\\
&&\qquad\qquad
\times g^{\mu\nu} B_0(p;M_\rho,M_{A_1})\,,
\nonumber\\
&&
\Pi_{\bar{\cal A}_{M}\bar{\cal A}_\perp}^{{\rm (f)}\mu\nu}
= \frac{N_f}{2} \Bigl(\frac{F}{F_\sigma}\Bigr)^2
  b g^{\mu\nu} A_0(M_\rho)\,,
\nonumber\\
&&
\Pi_{\bar{\cal A}_{M}\bar{\cal A}_\perp}^{{\rm (g)}\mu\nu}
= \frac{N_f}{2} \Bigl(\frac{F}{F_\pi}\Bigr)^2
  [\,2a(1-\zeta) - b(1-\zeta^2) - 2d\,] 
\nonumber\\
&&\qquad\qquad
\times g^{\mu\nu} A_0(0)\,,
\nonumber\\
&&
\Pi_{\bar{\cal A}_{M}\bar{\cal A}_\perp}^{{\rm (h)}\mu\nu}
= \frac{N_f}{2} \Bigl(\frac{F}{F_q}\Bigr)^2
  b g^{\mu\nu} A_0(M_{A_1})\,.
\label{eq:appB:4}
\end{eqnarray}
The quadratic and logarithmic divergences generated by
those diagrams are renormalized following 
Eq.~(\ref{renormalization}).

The renormalization group equations (RGEs) for the leading
order parameters ($a, b, c, d$ and $g$) take the
following forms:
\begin{eqnarray}
&&
\mu\frac{d (a F^2)}{d \mu}
= \frac{N_f}{(4\pi)^2}
\Biggl[ \mu^2\Bigl{\{} 
 \frac{1}{2} 
  {}+ \frac{1}{2}\Bigl(\frac{F}{F_\pi}\Bigr)^4 a^2 (1-\zeta^2)^2
\nonumber\\
&&
  {}+ \frac{1}{2}\Bigl(\frac{F}{F_q}\Bigr)^4 [ a - 2(b+c) ]^2
  {}+ \Bigl(\frac{F^2}{F_\pi F_q}\Bigr)^2 ( a\zeta - b )^2
\nonumber\\
&&
  {}+ 2\Bigl(\frac{F}{F_\pi}\Bigr)^2 \zeta( a\zeta - b )
  {}+ 2\Bigl(\frac{F}{F_q}\Bigr)^2 (a-b-c)
 \Bigr{\}}
\nonumber\\
&&
{}+
  \frac{3}{2}M_\rho^2 + 2\Bigl(\frac{F}{F_q}\Bigr)^2 a M_\rho^2
  {}+ 2\Bigl(\frac{F}{F_\pi}\Bigr)^2 g^2 F^2 ( a\zeta - b )^2
\nonumber\\
&&
  {}- 2\Bigl(\frac{F}{F_q}\Bigr)^2 (a-b-c) M_{A_1}^2
  {}- \frac{1}{2}\Bigl(\frac{F}{F_q}\Bigr)^4 [ a - 2(b+c) ]^2 M_{A_1}^2
\nonumber\\
&&
  {}- \frac{1}{2}\Bigl(\frac{F^2}{F_\pi F_q}\Bigr)^2
      ( a\zeta - b )^2 M_{A_1}^2
\Biggr]\,,
\label{eq:appB:5}
\end{eqnarray}
\begin{eqnarray}
&&
\mu\frac{d (b F^2)}{d \mu}
= \frac{N_f}{(4\pi)^2}
\Biggl[ \mu^2\Bigl{\{} 
  \Bigl(\frac{F}{F_\sigma}\Bigr)^2 b
\nonumber\\
&&
  {}+ 2\Bigl(\frac{F}{F_\pi}\Bigr)^2 [ - a\zeta 
    {}+ b\bigl( 1 - \frac{1}{2} (1-\zeta^2) \bigr) ]
\nonumber\\
&&
  {}+ \Bigl(\frac{F^2}{F_\sigma F_\pi}\Bigr)^2
    ( a - b\zeta ) ( a\zeta - b )
 \Bigr{\}}
\nonumber\\
&&
{}+
  \Bigl(\frac{F}{F_\sigma}\Bigr)^2 b M_\rho^2
{}- 2\Bigl(\frac{F}{F_q}\Bigr)^2 b 
     ( M_\rho^2 - \frac{3}{2}M_{A_1}^2 )
\nonumber\\
&&
  {}+ 2\Bigl(\frac{F}{F_\pi}\Bigr)^2 \zeta( a - b\zeta )
     ( M_\rho^2 - M_{A_1}^2 )
\nonumber\\
&&
  {}- \frac{1}{2}\Bigl(\frac{F^2}{F_\sigma F_\pi}\Bigr)^2
     ( a - b\zeta ) ( a\zeta - b ) M_\rho^2
\nonumber\\
&&
  {}+ \frac{1}{2}\Bigl(\frac{F^2}{F_\sigma F_q}\Bigr)^2
     ab ( M_\rho^2 + M_{A_1}^2 )
\Biggr]\,,
\label{eq:appB:6}
\end{eqnarray}
\begin{eqnarray}
&&
\mu\frac{d (c F^2)}{d \mu}
= \frac{N_f}{(4\pi)^2}
\Biggl[ \mu^2\Bigl{\{} 
   - \Bigl(\frac{F}{F_\sigma}\Bigr)^2 b
\nonumber\\
&&
  {}- 2\Bigl(\frac{F}{F_\pi}\Bigr)^2
   \bigl[ - a\zeta(1-\zeta) + \frac{1}{2}b(1-\zeta^2) - c\zeta^2 \bigr]
\nonumber\\
&&
  {}- \Bigl(\frac{F}{F_q}\Bigr)^2 (a-b-2c)
  {}+ \Bigl(\frac{F^2}{F_\sigma F_q}\Bigr)^2 ab
\nonumber\\
&&
  {}- \Bigl(\frac{F^2}{F_\sigma F_\pi}\Bigr)^2
     (a+b) (1-\zeta) ( a\zeta - b )
 \Bigr{\}}
\nonumber\\
&&
{}+
  \Bigl(\frac{F}{F_\sigma}\Bigr)^2 (2a - b) M_\rho^2
\nonumber\\
&&
  {}- 2\Bigl(\frac{F}{F_\pi}\Bigr)^2 
    \zeta [ a(1-\zeta) + c\zeta ] ( M_\rho^2 - M_{A_1}^2 )
\nonumber\\
&&
  {}+ 2\Bigl(\frac{F}{F_q}\Bigr)^2
    \Bigl[ \Bigl(\frac{3}{4}a - b - 2c \Bigr)M_\rho^2
\nonumber\\
&&\qquad\qquad
    {}+ \Bigl(-\frac{5}{4}a + \frac{3}{2}b + 3c \Bigr)M_{A_1}^2 \Bigr]
\nonumber\\
&&
  {}+ \frac{1}{2}\Bigl(\frac{F^2}{F_\sigma F_\pi}\Bigr)^2
    (a+b) (1-\zeta) ( a\zeta - b ) M_\rho^2
\nonumber\\
&&
  {}- \frac{1}{2}\Bigl(\frac{F^2}{F_\sigma F_q}\Bigr)^2
    ab (M_\rho^2 + M_{A_1}^2)
\Biggr]\,,
\label{eq:appB:7}
\end{eqnarray}
\begin{eqnarray}
&&
\mu\frac{d (d F^2)}{d \mu}
= \frac{N_f}{(4\pi)^2}
\Biggl[ \mu^2\Bigl{\{} 
  - \Bigl(\frac{F}{F_\sigma}\Bigr)^2 b
\nonumber\\
&&
  {}- \Bigl(\frac{F}{F_\pi}\Bigr)^2 [ 2a(1-\zeta) - b(1-\zeta^2) - 2d ]
\nonumber\\
&&
  {}- \Bigl(\frac{F}{F_q}\Bigr)^2 b
  {}+ \Bigl(\frac{F^2}{F_\sigma F_\pi}\Bigr)^2 (a+b) (1-\zeta)
    ( a - b\zeta )
\nonumber\\
&&
  {}+ \Bigl(\frac{F^2}{F_\sigma F_q}\Bigr)^2 b (a+b)
 \Bigr{\}}
{}-
  2\Bigl(\frac{F}{F_\sigma}\Bigr)^2 g^2 F^2 
   b \bigl( \frac{1}{2}a - b \bigr)
\nonumber\\
&&
  {}+ 2\Bigl(\frac{F}{F_\pi}\Bigr)^2 g^2 F^2
    ( a - b\zeta ) [ a(1-\zeta) + c\zeta ]
\nonumber\\
&&
  {}+ 2\Bigl(\frac{F}{F_q}\Bigr)^2 g^2 F^2
    b \bigl( a - \frac{1}{2}b - \frac{3}{2}c \bigr)
\nonumber\\
&&
  {}- \frac{1}{2}\Bigl(\frac{F^2}{F_\sigma F_\pi}\Bigr)^2
    (a+b) (1-\zeta) ( a - b\zeta )M_\rho^2
\nonumber\\
&&
  {}- \frac{1}{2}\Bigl(\frac{F^2}{F_\sigma F_q}\Bigr)^2
    b (a+b) (M_\rho^2 + M_{A_1}^2)
\Biggr]\,,
\label{eq:appB:8}
\end{eqnarray}
\begin{eqnarray}
&&
\mu\frac{d g^2}{d \mu}
= \frac{N_f}{(4\pi)^2}\frac{1}{12}
\Biggl[ 
 - \frac{351}{2} + \Bigl(\frac{F}{F_q}\Bigr)^2 a
\nonumber\\
&&
 {}+ \frac{1}{2}\Bigl(\frac{F}{F_\pi}\Bigr)^4 a^2 (1-\zeta^2)^2
 {}+ \frac{1}{2}\Bigl(\frac{F}{F_q}\Bigr)^4 [ a - 2(b+c) ]^2
\nonumber\\
&&
 {}+ \Bigl(\frac{F^2}{F_\sigma F_\pi}\Bigr)^2 ( a\zeta - b )^2
 {}+ \Bigl(\frac{F^2}{F_\pi F_q}\Bigr)^2 ( a\zeta - b )^2
\Biggr] g^4\,,
\nonumber\\
\label{eq:appB:9}
\end{eqnarray}
where $\mu$ denotes a renormalization scale.

When we take the parameter choice satisfying
$a=b$ and $d=0$, the RGEs are reduced to
\begin{eqnarray}
\mu\frac{d (a F^2)}{d \mu}
&=& \frac{N_f}{(4\pi)^2}\Bigl[ \mu^2 + 3 a g^2 F^2 \Bigr]\,,
\nonumber\\
\mu\frac{d (b F^2)}{d \mu}
&=& \frac{N_f}{(4\pi)^2}\Bigl[ \mu^2 + 3 a g^2 F^2 \Bigr]\,,
\nonumber\\
\mu\frac{d (c F^2)}{d \mu}
&=& \frac{N_f}{(4\pi)^2}\Bigl[ 2\mu^2 + 6c g^2 F^2 \Bigr]\,,
\nonumber\\
\mu\frac{d (d F^2)}{d \mu}
&=& 0\,,
\nonumber\\
\mu\frac{d g^2}{d \mu}
&=& - \frac{N_f}{(4\pi)^2}\frac{43}{3}g^4\,.
\end{eqnarray}


\end{document}